\documentclass[a4paper, 10pt]{article} %One column style
\usepackage{geometry}
\usepackage{titlesec}
\titleformat{\section}{\normalfont\large\bfseries}{\thesection}{1em}{}
\titleformat{\subsection}{\normalfont\normalsize\bfseries}{\thesubsection}{1em}{}

\titlespacing\section{0pt}{12pt plus 3pt minus 3pt}{3pt plus 1pt minus 1pt}
\titlespacing\subsection{0pt}{10pt plus 3pt minus 3pt}{3pt plus 1pt minus 1pt}
\titlespacing\subsubsection{0pt}{8pt plus 3pt minus 3pt}{3pt plus 1pt minus 1pt}

\def\abstract{\flushleft{\bf Abstract.}  } % refine abstract styling
\usepackage[numbers,sort&compress]{natbib}
\usepackage{microtype}
\usepackage{mathtools,amsfonts,dsfont,amsmath,amssymb}
\usepackage{xcolor}	    % colors for hyperlinks
\definecolor{mycolor}{RGB}{74,168,142}
\definecolor{mycolor}{RGB}{100,100,100}
\usepackage[colorlinks=true, linkcolor=mycolor, urlcolor=mycolor, citecolor=mycolor, anchorcolor=mycolor, backref=page]{hyperref}
\usepackage{amsmath}    % for align, etc.
\usepackage{mathtools}  % for bmatrix 
\usepackage{subcaption} % for subfigures
\usepackage[ruled,vlined]{algorithm2e} % for algorithms
\usepackage{todonotes}
\newcommand\blfootnote[1]{% A footnote without a marker
	\begingroup
	\renewcommand\thefootnote{}\footnote{#1}%
	\addtocounter{footnote}{-1}%
	\endgroup
}

\usepackage[markup=underlined]{changes}
\definechangesauthor[color=orange]{DG}

\definechangesauthor[color=blue]{WB}

\definechangesauthor[color=red]{JH}

\makeatletter
\renewcommand{\todo}[2][]{\@todo[caption={#2},size=\tiny, #1]{#2}}
\makeatother
\begin{document}   

%\title[Working title]{...}
\title{\vspace{-0.25cm}\Large \bf Deterministic Model of Incremental Multi-Agent Boltzmann Q-Learning: Transient Cooperation, Metastability, and Oscillations}

\author{David Goll$^\text{a}$, Jobst Heitzig$^\text{b}$ and Wolfram Barfuss$^\text{c}$} \date{} 
\maketitle
\thispagestyle{empty}

\begin{abstract}
	% ---------- ABSTRACT ---------- 
Multi-Agent Reinforcement Learning involves interacting agents whose learning processes are coupled through their shared environment, giving rise to emergent, collective dynamics that are sensitive to initial conditions and parameter variations.
A Dynamical Systems approach, which studies the evolution of multi-component systems over time, has uncovered some of the underlying dynamics by constructing deterministic approximation models of stochastic algorithms.
In this work, we demonstrate that even in the simplest case of independent Q-learning with a Boltzmann exploration policy, significant discrepancies arise between the actual algorithm and previous approximations. 
We elaborate why these models actually approximate interesting variants, simplifying the learning dynamics, rather than the original incremental algorithm. 
To explain the discrepancies, we introduce a new discrete-time approximation model that explicitly accounts for agents' update frequencies within the learning process, and show that its dynamics fundamentally differ from the simplified dynamics of prior models.
We illustrate the usefulness of our approach by applying it to the question of spontaneous cooperation in social dilemmas, specifically the Prisoner's Dilemma as the simplest case study. 
We identify conditions under which the learning behaviour appears as long-term stable cooperation from an external perspective.
However, our model shows that this behaviour is merely a metastable transient phase and not a true equilibrium, making it exploitable.
We further exemplify how specific parameter settings can significantly exacerbate the moving target problem in independent learning.
Through a systematic analysis of our model, we show that increasing the discount factor induces oscillations, preventing convergence to a joint policy. These oscillations arise from a supercritical Neimark-Sacker bifurcation, which transforms the unique stable fixed point into an unstable focus surrounded by a stable limit cycle.

\end{abstract}
\vspace{-0.125cm}

\blfootnote{\!\!\!\!\!\!\!\!\!\!\!
\textbf{Preprint working paper.} \today. Feedback welcome. \newline
% \textbf{Manuscript for review.} \today. \newline
	$^\text{a}$\! Humboldt University of Berlin;
	$^\text{b}$\! Potsdam Institute of Climate Impact Research;
	$^\text{c}$\! University of Bonn;\\
	\textbf{Contact}: david.ben.goll@gmail.com, wbarfuss@uni-bonn.de}

% keywords should be added
% ...
% --------------- 

%---------------------- Main Body ----------------------
\section{Introduction}
Reinforcement Learning (RL) \cite{Sutton1998} is a foundational machine learning approach where a singular agent learns optimal behaviours through trial-and-error interactions with its environment to maximise expected cumulative rewards. 
A classic method in RL is tabular Temporal-Difference (TD) learning, with Q-learning (QL)---hereafter referring specifically to the original incremental algorithm introduced by Watkins \cite{watkins1992}---being one of its most prominent examples.
Interestingly, TD learning closely resembles biological learning mechanisms, as studies show that dopamine signals in the brain encode reward-prediction errors, highlighting parallels between artificial and biological learning \cite{schultz1997neural, dayan2008reinforcement, doi:10.1126/science.275.5306.1593}.

While single-agent RL has established a strong theoretical foundation and has been successfully applied in many scenarios \cite{mnih2013playing, mnih2015human, silver2016mastering}, real-world problems often involve multiple agents interacting in shared environments \cite{dafoe2020open, dafoe2021cooperative}. Multi-Agent Reinforcement Learning (MARL) \cite{marl-book} extends RL to these settings, where agents learn concurrently while adapting to each other's behaviours. 
The independent learning approach \cite{Tan-Ming} naturally extends single-agent algorithms to multi-agent settings by treating each agent's learning as an isolated process. The processes are `independent'' insofar as they are only indirectly linked to other agents' learning processes via the shared environment. 
However, this extension results in a significant complication: the loss of convergence guarantees that are present in single-agent settings if the environment is stationary \cite{hernandez2017survey}. 
In theoretical studies of RL, the environment is typically modelled as a Markov Decision Process (MDP), described in terms of states, stationary transition probabilities, and rewards.
In the independent learning approach, each agent perceives the evolving policies of other agents as part of its ``effective" environment, rendering it non-stationary.
The independent learning approach is thus incompatible with the MDP framework on which most convergence guarantees in single-agent algorithms rely.
Despite this lack of convergence guarantees, independent learning is often used in practice due to its adaptability and scalability with the number of agents \cite{matignon2012independent, hernandez2019survey}. These simple algorithms serve as crucial baselines in MARL research and can achieve individual cumulative rewards that are comparable with more sophisticated state-of-the-art methods \cite{marl-book, papoudakis2020benchmarking}.

Irrespective of the details of the collective learning process in MARL, the interdependence of agents and the inherent stochasticity of the algorithms often lead to complex emergent dynamics. 
To enable a detailed analysis of these dynamics, a complementary perspective from Dynamical Systems theory has proven beneficial \cite{BORGERS19971, tuyls_2003, Farmer_complexChaos2P_2002, Sato_2003, Sato_2005, leslie2005individual, izquierdo2008reinforcement, fudenberg2009learning, Galla_2009, masuda2009theoretical, wunder2010classes, masuda2011numerical, Galla_2011, Kianercy_2012, galstyan2013continuous, Farmer_Galla_2013, bloembergen2015evolutionary, FarmerGallaSanders2018, Barfuss2019, hu2019modelling, Barfuss21, hu2022dynamics, chu2022formal, barfuss2022modeling, leonardos2022exploration}.
To this end, the stochastic algorithms are approximated in deterministic dynamical equations, which allows for a concise interpretation of the underlying dynamics and provides a framework for a rigorous analysis of the effects of parameters and initial conditions. 
In particular, a common approach is to take a continuous-time limit and link the resulting ordinary differential equations to the replicator dynamics of Evolutionary Game Theory \cite{BORGERS19971, Farmer_complexChaos2P_2002, Sato_2003, tuyls_2003, Galla_2009}. 
For the specific algorithm of independent Q-learning in a single-state environment, this was initially done in \citeyear{tuyls_2003} by \citeauthor{tuyls_2003} \cite{tuyls_2003}. 
However as pointed out by \citeauthor{kaisers2010frequency} in \citeyear{kaisers2010frequency}, the procedure applied in \cite{tuyls_2003} actually approximates a modified variant of the algorithm \cite{kaisers2010frequency} resulting in discrepancies in the learning dynamics.
While some publications \cite{leslie2005individual, kaisers2010frequency, bloembergen2015evolutionary, hernandez2017survey, Barfuss21} acknowledged that the model does not represent independent Q-learning, others---including \cite{Kianercy_2012}, \cite{galstyan2013continuous}, and more recent works like \cite{leonardos2022exploration} and \cite{mintz2024evolutionary}---do not mention this discrepancy, potentially overlooking its implications.
Noteworthy, in \citeyear{Barfuss2019}, \citeauthor{Barfuss2019} extended the method of \cite{tuyls_2003} to multi-state environments, by separating the interaction and adaptation timescales. 
In the case of single-state environments, the resulting deterministic learning dynamics correspond to a discrete-time version of the initial version derived in \cite{tuyls_2003}.

In this work, we clarify the relationships between both previous models and the original algorithm and elaborate why they cannot describe the stochastic learning dynamics of independent Q-learning.
We then propose an alternative approximation model, which explains why agents appear to ``learn" to spontaneously cooperate over extended periods in social dilemmas and shows that this behaviour is not a true equilibrium but merely a metastable phase of the dynamics.

The paper is organised as follows: In section \ref{sec:Background}, we introduce the independent Q-learning algorithm and the specific setup under study. We then review the two aforementioned existing models of \cite{tuyls_2003} and \cite{Barfuss2019}, and elaborate why they actually approximate interesting variants of independent Q-learning, rather than describing the fundamentally more complex dynamics of the original algorithm. To highlight these stylised discrepancies, we compare the deterministic dynamics with the actual stochastic learning process in the context of the Prisoner's Dilemma.
In section \ref{sec:ourModel}, we propose an alternative deterministic approximation model for single-state environments and demonstrate its effectiveness in capturing the emergent behaviour of independent Q-learning. 
Our model enables us to distinguish between metastable phases and true equilibria and explains how stable oscillations arise from the moving-target problem, preventing convergence under certain parameter settings, even in simple setups such as the Prisoner's Dilemma.
We conclude with a discussion of the broader implications of our findings, emphasising the limitations of the independent learning approach and the need for caution when interpreting results.
\section{Background, Problem and Pitfalls} \label{sec:Background}

% -------------- Subsection --------------
\subsection{Independent Q-Learning in a Single-State Environment}\label{sec:IndependentQLearning}
We study the simplest possible multi-agent system: two agents interacting in a single-state environment, playing the Prisoner's Dilemma as the most paradigmatic example game. The Prisoner's Dilemma, characterised by a single Nash equilibrium where both agents defect, along with its iterated version, has been extensively studied in the MARL community \cite{SANDHOLM1996147, tuyls_2003, izquierdo2008reinforcement, masuda2009theoretical, masuda2011numerical, kaisers2010frequency, kaisers2011faq, wang2022levy, meylahn2022limiting, bertrand2023q, usui2021symmetric}.
The reward tensor for the game is given by
\begin{equation}
    \label{eq:Prisoner'sDilemma}
        \mathbf{R} 
        =
        \left (
            \begin{array}{cc}
            R_{CC}^{1}, R_{CC}^{2} & R_{CD}^{1}, R_{CD}^{2} \\
            R_{DC}^{1}, R_{DC}^{2} & R_{DD}^{1}, R_{DD}^{2}
        \end{array}
        \right )
        = 
        \left(
        \begin{array}{cc}
        3, 3 & 0, 5 \\
        5, 0 & 1, 1
        \end{array}
        \right)
        .
\end{equation}
The agents can either choose to cooperate ($C$) or to defect ($D$).
Throughout this work, the superscript index $i$ denotes an agent, while $-i$ represents its opponent. Random variables are denoted by uppercase letters (e.g. $A^i$), their specific instances in lowercase (e.g. $a^i$), and tensors in boldface (e.g. $\mathbf A$).
At each time step $t$, each agent $i$ chooses an action $A^i(t) = a^i \in \mathcal A^i$, where $\mathcal A^i = \{ C, D \}$,
receives a reward $R^i_{\mathbf A (t)}$ based on the joint action $\mathbf A (t) = (A^i(t), A^{-i}(t)) \in \mathcal A$, where $\mathcal A = \mathcal A^1 \times \mathcal A^2$, and updates its policy accordingly. The process repeats until a terminal time step is reached.

The agents adapt to new information via independent Q-learning (algorithm \ref{algorithm IQL}).
The state-action value estimate $Q^i_{a^i}(t)$, called $Q$-value, represents how much agent $i$ values action $a^i$ at time $t$.
The stochastic update rule reads
\begin{align}
\label{eq:StochasticQ-UpdateRule}
    Q^i_{a^i}(t+1) 
    &= 
    Q^i_{a^i}(t) + 
    \alpha^i
    \delta_{A^i(t) a^i}
    \left[ R^i_{ \mathbf{A} (t) }
    + \gamma^i \max_{b^i \in \mathcal A^i} Q^i_{b^i}(t) 
    - Q^i_{a^i}(t) \right]
    ,
\end{align}
where $\alpha^i \in [0,1)$ is called the agent's \textit{learning rate} and $\gamma^i \in [0,1)$ its \textit{discount factor}. 
The policy $\pi^i_{a^i}(t)$ is the probability of agent $i$ to choose action $a^i$ at time $t$. It is drawn from a Boltzmann distribution
\begin{align}
\label{eq:BoltzmannPolicy}
    \pi^i_{a^i}(t) &:= f(Q^i(t), a^i) ,
    \\
    f(Q^i, a^i) 
    &:= 
    \frac{ \exp[ Q^i_{a^i} / T^i ] }
    {\sum_{b^i \in \mathcal A^i} \exp[ Q^i_{b^i}  / T^i] } 
    .
\end{align}
where $T^i \in (0, \infty)$ is called \textit{temperature} in analogy to statistical physics. Before we continue, we want to remark some comments on this setup and why it is of interest.

% ------ ALGORITHM BEGINNING ------
\begin{algorithm}[t]
\caption{Independent Q-learning with Boltzmann policy in a single-state environment}
\label{algorithm IQL}
\SetAlgoLined
\KwIn{Action space $\mathcal{A}^i$, learning rate $\alpha^i$, discount factor $\gamma^i$, temperature parameter $T^i$ for each agent $i$, common environment E}
\KwOut{Learned $Q$-values $Q^i_{a^i}$ for each agent $i$}
% Initialize $Q^i_{a^i}$ arbitrarily for all $s \in \mathcal{S}, a \in \mathcal{A}^i$ and for each agent $i$;
Initialise $Q^i_{a^i}$ arbitrarily for all $a^i \in \mathcal{A}^i$ for each agent $i$ \\
    \While{not reached terminal time step}{ \
        \For{each agent $i$}{
            Choose action $a^i$ with Boltzmann policy: \\
            \Indp
            $\pi^i_{a^i} \gets 
            \displaystyle\frac{e^{Q^i_{a^i} / T^i}}{\sum_{b^i} e^{Q^i_{b^i}  / T^i}}$ for all $a^i \in \mathcal{A}^i$ 
            \\
            $a^i \sim \pi^i_{a^i}$ \\
            \Indm
        }
        Take joint action $\mathbf{a} = (a^1, a^2, ..., a^n)$ in the environment E\\
        \For{each agent $i$}{
            Observe own reward $r^i$ in the environment E \\
            Update $Q$-value of chosen action: \\
            \Indp
            $\displaystyle Q^i_{a^i} \leftarrow Q^i_{a^i} + \alpha^i \left[ r^i + \gamma^i \max_{b^i} Q^i_{b^i}  - Q^i_{a^i} \right]$
            \Indm
        }
    }

\end{algorithm}
% ------ ALGORITHM END ------

% comments:
    Formally, the environment consists of a single \emph{non-terminal} state, $s_0$, defined by \eqref{eq:Prisoner'sDilemma}. After each time step, the environment transitions back to $s_0$. The learning process concludes at a terminal time step, at which point the environment transitions to a \emph{terminal} state, $s_{terminal}$. By definition, no rewards are provided in the terminal state, and agents remain there indefinitely \cite{marl-book}. 
    We note that this setup corresponds to the game-theoretic definition of a finitely repeated normal-form game, where agents do \emph{not} condition their policies on past interactions.
    
    In the original single-agent Q-learning algorithm (see Appendix \ref{sec:Appendix_QL_original} or \cite{watkins1992}), the discount factor $\gamma^i$ is a hyperparameter that determines an agent's preference for future state values in \emph{multi-state} environments. 
    The necessity of including a discount factor in a \emph{single-state} environment, as considered here, is therefore debatable.
    Some studies effectively set $\gamma^i = 0$ by defining the environment to transition into a terminal state after each round \cite{galstyan2013continuous, Kianercy_2012, leonardos2022exploration, hu2022dynamics}. Others define the environment as static yet repetitive and keep the term involving $\gamma$ \cite{tuyls_2003, babes2009q, wunder2010classes, kaisers2010frequency, zschache2018melioration, mintz2024evolutionary}. 
    To preserve the algorithm's core structure---where the term involving $\gamma^i$ is a defining feature---we consider a repetitive environment and retain the discount factor. 
    Given that the agents lack knowledge of when the game will end, our framework is consistent with the common interpretation of $\gamma^i$ to be the agent's belief about the probability that the game continues in the next time step.

    We adopt the term `policy' ($\pi$) to describe agents' action probabilities to stay consistent with machine learning conventions \cite{Sutton1998, marl-book}. The Boltzmann policy function is chosen over common alternatives like epsilon-greedy because it uses a smooth probability distribution based on $Q$-values rather than discrete choices. 
    Some studies suggest this mechanism aligns with human and animal decision-making in competitive and observational learning tasks \cite{lee2004reinforcement, kim2009valuation}. %burke2010neural 
    The temperature parameter $T^i > 0$ regulates the exploration-exploitation trade-off: higher $T^i$ promotes exploration by equalising probabilities, while lower $T^i$ emphasises exploitation of actions with higher $Q$-values. As $T^i \to 0$, the agent converges to a pure policy. We keep the temperature constant throughout the learning process, rather than annealing it \cite{SANDHOLM1996147}, to simplify the process and enhance the interpretability of the results.

    In RL, the outcome of the learning process is typically interpreted as a pure policy: the action with the maximum Q-value in a given state is regarded as the ``learned" action. However, in this work, we focus on the dynamics of the learning process itself, interpreting the Boltzmann distribution as the ``learned" policy at any time $t$, as it reflects the agent's probabilistic decision-making process. Our primary interest lies in understanding the long-term behaviour of the learning process as a function of parameters and initial conditions. 
    
    The dynamics of the system are fully described by the four-dimensional state vector in Q-space, $\mathbf{Q}(t) := (Q^1_C(t), Q^1_D(t), Q^2_C(t), Q^2_D(t))$. The four $Q$-values are the fundamental dynamical variables, evolving according to \eqref{eq:StochasticQ-UpdateRule}.  
    In contrast, the joint policy $\boldsymbol \pi (t)$ resides in a two-dimensional subspace due to the normalisation constraint, $\sum_{a^i \in \mathcal{A}^i} \pi^i_{a^i}(t) = 1$, for all $i$ and $t$. 
    At any time $t$, this subspace is represented by $\boldsymbol{\pi}_C(t) = (\pi^1_C(t), \pi^2_C(t))$, capturing agents' cooperation probabilities. 
    Thus, the Q-space encodes the full state of the system, while the policy space offers a lower-dimensional representation, indicating what the agents will actually do.

    It is important to note that the use of the Kronecker delta $\delta_{A^i(t) a^i}$ in the update rule \eqref{eq:StochasticQ-UpdateRule} implies that \emph{only} the $Q$-value of the action $A^i(t)$ played at time $t$ by agent $i$ is updated, while the remaining $Q$-values retain their current values. As will be shown in the next section, this feature is critical to the algorithm’s structure; neglecting or modifying it results in dynamics that diverge from the original formulation \cite{watkins1992}, that is commonly used \cite{Sutton1998, marl-book}.

% -------------- Subsection --------------
\subsection{Previous Deterministic Models}
The stochastic nature of MARL makes its dynamics obscure and difficult to interpret \cite{hernandez2017survey, hernandez2019survey}. 
The primary goal of constructing approximation models of MARL systems is thus to transform the algorithms into deterministic dynamical equations, either in discrete or continuous time, which enables a convenient analysis. 
A secondary goal may be to reduce the complexity of a system to its essentials. In the specific setup considered in this work, this could mean to reduce the dynamics from the four-dimensional Q-space into the two-dimensional policy space. 
Here, we present two existing approximation models of MARL that take this step and explain why, in doing so, they deviate from classic incremental Q-learning—defined by \eqref{eq:StochasticQ-UpdateRule}—and instead represent modified variants. 

% -------------- Subsubsection --------------
\paragraph{Frequency-Adjusted Q-learning (FAQL) Model} In \citeyear{BORGERS19971}, \citeauthor{BORGERS19971} first established a connection between MARL and the replicator dynamics of Evolutionary Game Theory by deriving a continuous-time limit for the \textit{Cross Learning} \cite{Cross1973} algorithm. 
The continuous-time limit is constructed by segmenting the time into intervals $\Delta t$, substituting the time step $(t + 1)$ with $(t + \Delta t)$, the learning rate $\alpha$ with $\alpha' = \alpha \Delta t $, and taking the limit $\Delta t \rightarrow 0$.

Building on this foundation, \citeauthor{tuyls_2003} in \citeyear{tuyls_2003}---and similarly \citeauthor{Sato_2003} for a slightly different variant---applied this approach to independent Q-learning in single-state environments. The authors proposed that the time evolution of an agent's Boltzmann policy in the continuous-time limit can be approximated by the deterministic replicator equation
\begin{equation}
\begin{aligned}
    \label{eq:FAQL_model}
    \frac{d}{dt} \pi^i_{a^i}(t) 
    &= 
    \frac{\alpha^i}{ T^i } \pi^i_{a^i}(t) 
    \left ( 
    \mathbb E_{A^{-i}(t) \sim \pi^{-i}(t)} R^i_{a^i A^{-i}(t)} 
    - \sum_{b^i \in \mathcal A^i} 
    \mathbb E_{A^{-i}(t) \sim \pi^{-i}(t)} R^i_{b^i A^{-i}(t)}
    \right )
    \\
    & \quad +
    \alpha^i \pi^i_{a^i}(t) 
    \sum_{b^i \in \mathcal A^i}
    \pi^i_{b^i}(t) 
    \ln \frac{\pi^i_{b^i}(t) }{\pi^i_{a^i}(t) }
    .
\end{aligned}
\end{equation}
But in the derivation, it was implicitly assumed\footnote{This non-trivial assumption is not explicitly stated.}  
that \emph{all} $Q$-values are updated at each time step, effectively treating the update rule 
\eqref{eq:StochasticQ-UpdateRule} as if the Kronecker delta $\delta_{A^i(t), a^i}$ were absent.
This assumption allows to simplify the dynamics to the lower-dimensional policy space and eliminates all terms involving the discount factor $\gamma^i$.
While this may aid theoretical analysis, it introduces significant discrepancies between the model and actual dynamics \cite{kaisers2010frequency}.

However, the model aligns well with a modified variant of Q-learning, termed `\emph{frequency-adjusted} Q-learning' (FAQL). 
Originally applied by \citeauthor{leslie2005individual} in \citeyear{leslie2005individual}, \citeauthor{kaisers2010frequency} defined and termed it a separate algorithm in \citeyear{kaisers2010frequency}, after identifying the discrepancies to be caused by the update frequencies.
But rather than revising the approximation model to reflect actual learning dynamics, they adjusted the algorithm itself to match the simplified model dynamics, arguing that this adaptation yields more favourable and stable outcomes. 
For clarity, we therefore refer to the simplified model \eqref{eq:FAQL_model} as the `FAQL model' throughout the remainder of this work. 

The FAQL algorithm \cite{kaisers2010frequency} smooths the learning process by scaling the learning rate with the inverse of the update frequency, $1 / \pi^i_{a^i}(t)$, which effectively diminishes the influence of the Kronecker delta in the derivation of \eqref{eq:FAQL_model}.
Additionally, it introduces a new hyperparameter, $\beta^i \in [0, 1)$, which modifies the update rule\footnote{The minimum ensures that the effective learning rate does not exceed one.} to
\begin{equation}
\label{eq:FAQ value update rule}
   Q^i_{a^i}(t+1) = 
   Q^i_{a^i}(t) + 
        \alpha^i 
        \min 
        \left(
        \frac{\beta^i}{\pi^i_{a^i}(t)}, 1
        \right)
        \delta_{A^i(t) a^i}
        \left[ 
        R^i_{\mathbf A (t)} + \gamma^i \max_{b^i \in \mathcal A^i} Q^i_{b^i}(t) - Q^i_{a^i}(t) 
        \right]
        .
\end{equation}

% -------------- Subsubsection --------------
\paragraph{Batch Q-Learning (BQL) Model}
In \citeyear{Barfuss2019}, \citeauthor{Barfuss2019} extended previous deterministic models of MARL, such as the FAQL model, which were focused so far on single-state environments, to encompass \emph{multi-state} environments  with time discounting \cite{Barfuss2019}.
While its title might suggest it represents classic, incremental Temporal Difference (TD) learning---of which Q-learning is a specific case---the model actually approximates a \emph{batch} version of TD-learning rather than incremental TD-learning. 
In batch learning \cite{BatchRL-chapter-2012}, the timescales of interaction and adaptation are separated. 
This approach allows agents to adapt based on aggregated experiences rather than individual interactions.
Here we follow the definition of \cite{Barfuss2019}, restrict it to the single-state setup as defined in section \ref{sec:IndependentQLearning} and discuss its deterministic approximation. 

The agents interact $K \in \mathbb N_+$ times under the \emph{constant} joint policy $\boldsymbol \pi (t)$. The information from these interactions are stored inside a batch of size $K$. At the update step $(t + K)$, the agents then use the sample average of the gathered experience to update their $Q$-values and subsequently the joint policy $\boldsymbol \pi (t + K)$. With a minor abuse of notation to improve readability, \eqref{eq:StochasticQ-UpdateRule} is modified to 
\begin{align}
    Q^i_{a^i}(t+K) 
    &= 
    Q^i_{a^i}(t) + 
    \alpha^i
    D^i_{a^i, \mathbf A(t), \dots ,\mathbf A(t+K), Q^i(t)}
    ,
    \\
    D^i_{a^i, \mathbf A(t), \dots ,\mathbf A(t+K), Q^i(t)} 
    &:= 
    \frac{1}{K_{a^i}} \sum_{k=0}^{K-1} 
        \delta_{A^i(t+k) a^i}
        \left[ 
            R^i_{\mathbf A (t+k)}
            + \gamma^i \max_{b^i \in \mathcal A^i} Q^i_{b^i}(t) 
            - Q^i_{a^i}(t) 
        \right]
    ,
    \label{eq:batchTemporalDifferenceError}
\end{align}
where $K_{a^i} := \max \left( 1 , \sum_{k=0}^{K-1} \delta_{A^i(t+k) a^i} \right)$ denotes the number of times agent $i$ played action $a^i$. To avoid division of zero, $K_{a^i} := 1$ if the action $a^i$ was never played. 
For a batch size of $K=1$, batch Q-learning is equal to regular Q-learning.
Note however that for $K>1$, batch learning allows to update \emph{multiple} $Q$-values per agent per update step---all $Q$-values whose actions were played in the batch. 

In the infinite batch limit $K \rightarrow \infty$ (and subsequently $K_{a^i} \rightarrow \infty$), the stochastic batch temporal difference error \eqref{eq:batchTemporalDifferenceError} becomes almost surely (a.s.) deterministic due to the law of large numbers. 
The limit implies that, with probability one, \emph{all} $Q$-values are updated simultaneously at each update step. 
This enables the derivation of a deterministic update rule in the separated \emph{update timescale $u$} that operates exclusively in the lower-dimensional policy space (see appendix \ref{sec:Appendix_BQL_derivation} for a detailed derivation):
\begin{equation}
\begin{aligned}
\label{eq:BQL_model}
    \pi^i_{a^i}(u + 1)
    &= \frac{
    \pi^i_{a^i}(u) 
    \exp[ \alpha^i
        D^i_{a^i \boldsymbol \pi(u)}  / T^i ]
    }{
    \sum_{b^i \in \mathcal A^i} \pi^i_{b^i}(u) 
    \exp[ \alpha^i  
    D^i_{b^i \boldsymbol \pi(u)}  / T^i ]
    }
    ,
\end{aligned}
\end{equation}
where
\begin{equation}
\begin{aligned}
\label{eq:BQL_model_ReducedDeterministicTDerror}
    D^i_{a^i, \boldsymbol \pi(u)} 
        &:=
        \mathbb E _{A^{-i}(u) \sim \pi^{-i}(u)}
        R^i_{a^i A^{-i}(u)} 
        - T^i \ln \pi^i_{a^i}(u)  
        .
\end{aligned}
\end{equation}
Note that for single-state environments, all terms which include the discount factor $\gamma^i$ vanish in the derivation of \eqref{eq:BQL_model}. As in the FAQL model, this is again due to the implicit assumption that all $Q$-values get updated simultaneously.
\citeauthor{Barfuss21} demonstrated good agreement of \eqref{eq:BQL_model} with actual behaviour for $K \approx 10^3 - 10^4$, but not for smaller $K$-values \cite{Barfuss21}. To emphasise its distinction from incremental Q-learning, we will refer to this model throughout this work as the `Batch Q-Learning' (BQL) model. In single-state environments, the FAQL model corresponds to the continuous-time limit of the BQL model; hence, we also collectively refer to them as the `FAQL/BQL model'.

A fixed point policy $\boldsymbol \pi_{*}$ of \eqref{eq:BQL_model} can be determined by finding the roots of \eqref{eq:BQL_model_ReducedDeterministicTDerror} for all $i, a^i$. 
After normalisation, this results in the two-dimensional system of equations
\begin{equation}
\label{eq:FixedPointPi}
    \pi^i_{a^i*} = \frac{
    \exp[ 
        \mathbb E _{A^{-i} \sim \pi^{-i}_*}  R^i_{a^i A^{-i}} /T  
        ]
    }{
    \sum_{b^i \in \mathcal A^i} 
    \exp[ 
        \mathbb E _{A^{-i} \sim \pi^{-i}_*}  R^i_{b^i A^{-i}} /T 
        ]
    } 
    .
\end{equation}
This equation can also be interpreted outside the learning context as defining a ``soft'' version of Nash equilibrium based on a form of bounded rationality rather than full rationality: if the equation is fulfilled, both players do not maximise but ``soft maximise'' their reward under the correct assumption that the other player does likewise, by playing the corresponding Boltzmann policy. In behavioural game theory, this form of equilibrium is called `Logit Quantal Response equilibrium' \cite{mckelvey1995quantal}. As experimental evidence from humans suggest that indeed boundedly rational human decisions sometimes approximate such soft equilibria \cite{teeselink2024high}, the question of whether MARL algorithms converge to such points as well is an important plausibility check. \\
\\
In summary, the dynamics of both previous approximation models, FAQL and BQL, exhibit the following key characteristics:
\begin{enumerate} 
    \item A fixed point of the dynamics is a boundedly rational strategic equilibrium.

    \item They are fully described within the lower-dimensional policy space.
    
    \item For single-state environments, they are independent of the discount factor $\gamma^i$---all terms including $\gamma^i$ vanish in the derivation. 

\end{enumerate}
In the following section, we compare them to stochastic realisations of independent Q-learning. We aim to answer whether the FAQL/BQL model still capture the core principles of Q-learning in a multi-agent setting, or if the inherent assumptions cause the approximated dynamics to deviate significantly from independent Q-learning.

% ----------- FIGURE 1 ------------
\begin{figure}[t]
% Instructions: \xdim and \ydim should be set such that the labels are in the top left corner of each subfigure!
\newcommand{\xdim}{-205}
\newcommand{\ydim}{160}
        \begin{subfigure}[t]{0.49\textwidth}
            \centering
            \includegraphics[width=\textwidth]{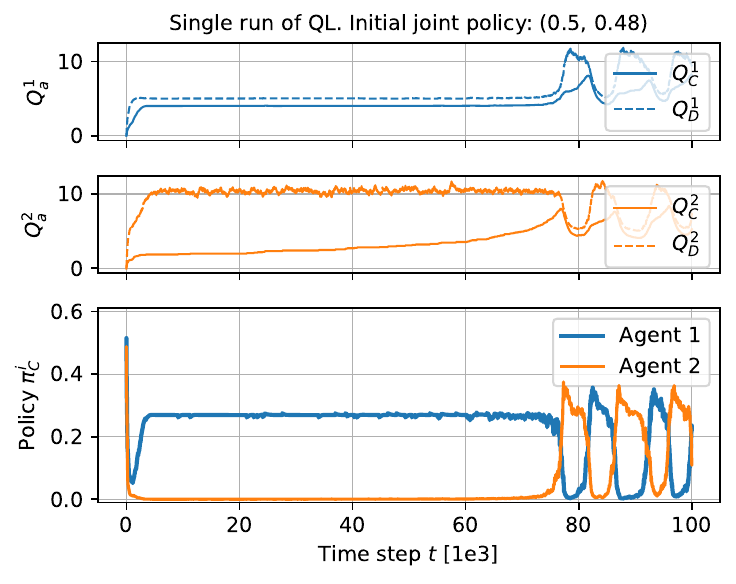}
            \put(\xdim, \ydim){A} % Top-left label
        \end{subfigure}
        \hfill
        \begin{subfigure}[t]{0.49\textwidth}
            \centering
            \includegraphics[width=\textwidth]{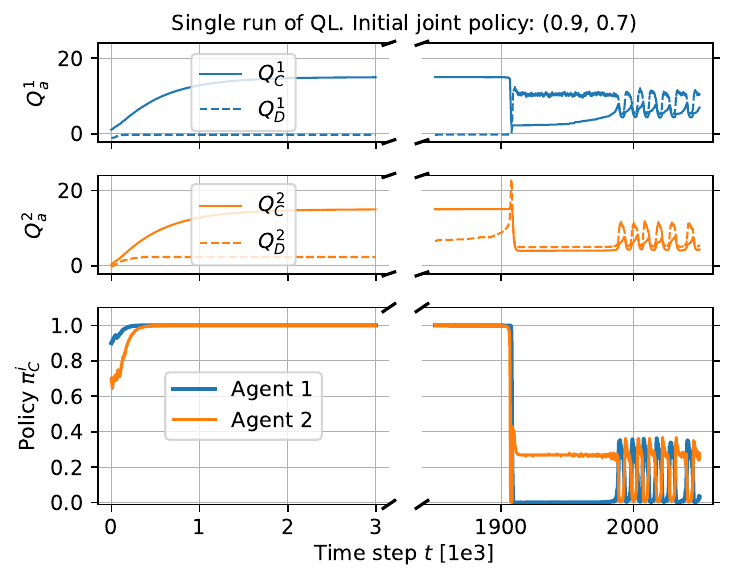}
            \put(\xdim, \ydim){B} % Top-left label
        \end{subfigure}
        \hfill
        \begin{subfigure}[t]{0.49\textwidth}
            \centering
            \includegraphics[width=\textwidth]{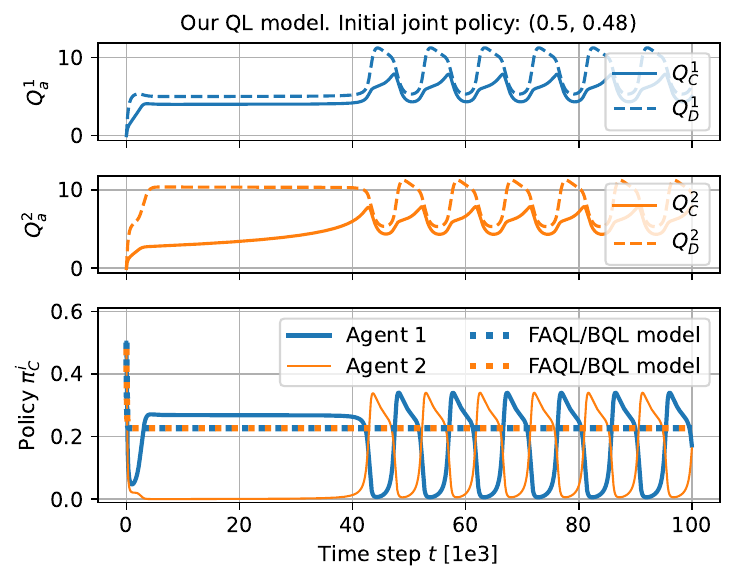}
            \put(\xdim, \ydim){C} % Top-left label
        \end{subfigure}
        \hfill
        \begin{subfigure}[t]{0.49\textwidth}
            \centering
            \includegraphics[width=\textwidth]{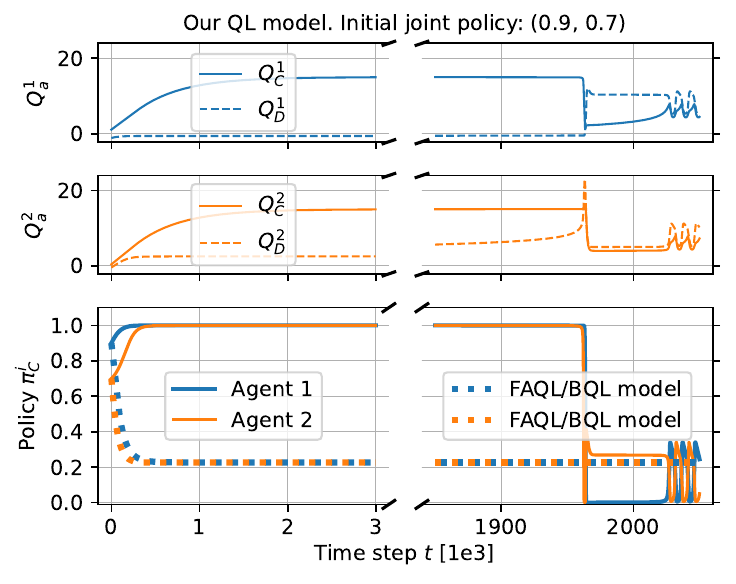}
            \put(\xdim, \ydim){D} % Top-left label
        \end{subfigure}
\caption{ 
Comparison between a single run of independent Q-learning on the Prisoner's Dilemma (top panels: A, B) and our deterministic approximation model (bottom panels: C, D), defined by \eqref{eq:QLmodelNEW}, for $T=1$, $\alpha = 0.01$, $\gamma = 0.8$, $Q_{base} = 0$. 
Note that the depicted runs in A and B represent single instances of a stochastic process. Timings and trajectories vary across different runs.
The first two subplots in each panel show the evolution of the $Q$-values ($Q^1_C, Q^1_D, Q^2_C, Q^2_D$), while the third subplot illustrates the resulting probabilities of cooperation ($\pi^1_C, \pi^2_C$).
The dotted policy trajectories in C and D represent previous approximation methods: FAQL, defined by \eqref{eq:FAQL_model}, and BQL, defined by \eqref{eq:BQL_model}.
The left panels (A, C) depict an initial joint policy $(\pi^1_C, \pi^2_C) = (0.5, 0.48)$, corresponding to $Q$-values $(0, 0, -0.04, 0.04)$ via \eqref{eq:Q-value_initialisation}. The right panels (B, C) show an initial joint policy $(\pi^1_C, \pi^2_C) = (0.9, 0.7)$, corresponding to $Q$-values $(1.1, -1.1, 0.4, -0.4)$ via \eqref{eq:Q-value_initialisation}.
}
\label{fig:1}
\end{figure}
% -----------

% ----------- FIGURE 2 ------------
\begin{figure}
% QL DYNAMICS
% Instructions: \xdim and \ydim should be set such that the labels are in the top left corner of each subfigure!
\newcommand{\xdimI}{-205}
\newcommand{\ydimI}{170}
    % QL DYNAMICS for Q-base = 0
    \centering
    \begin{subfigure}[b]{0.49\textwidth}
        \centering
        \includegraphics[width=\textwidth]{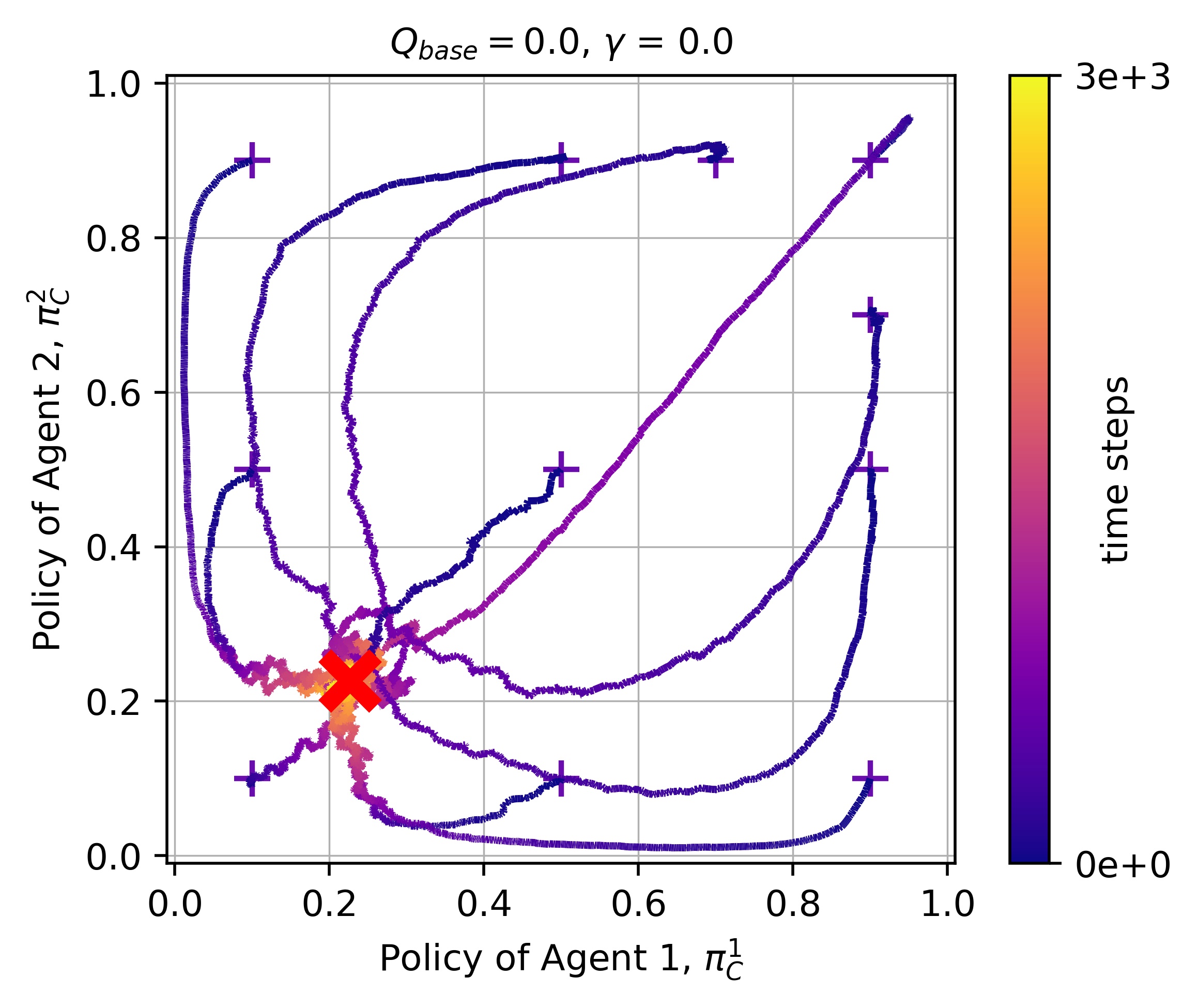}
        \put(\xdimI,185){\large \textbf{I}} % Top-left label for FIGURE I. Should be at the top left corner of the first row
        \put(\xdimI, \ydimI){A} % Top-left label
    \end{subfigure}
    \hfill
    \begin{subfigure}[b]{0.49\textwidth}
        \centering
        \includegraphics[width=\textwidth]{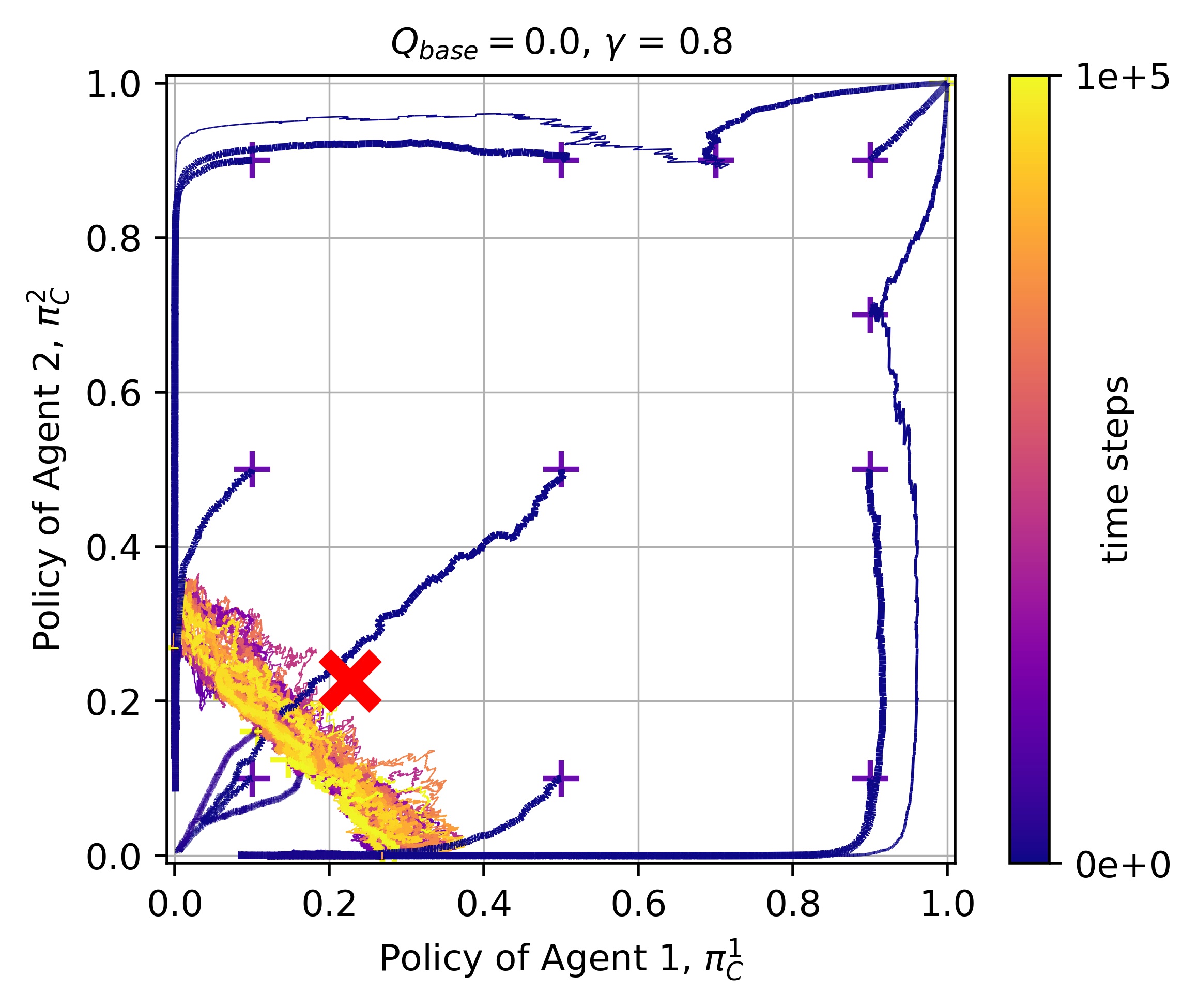}
        \put(\xdimI, \ydimI){B} % Top-left label
    \end{subfigure}
    % QL DYNAMICS for Q-base = 5
    \centering
    \begin{subfigure}[b]{0.49\textwidth}
        \centering
        \includegraphics[width=\textwidth]{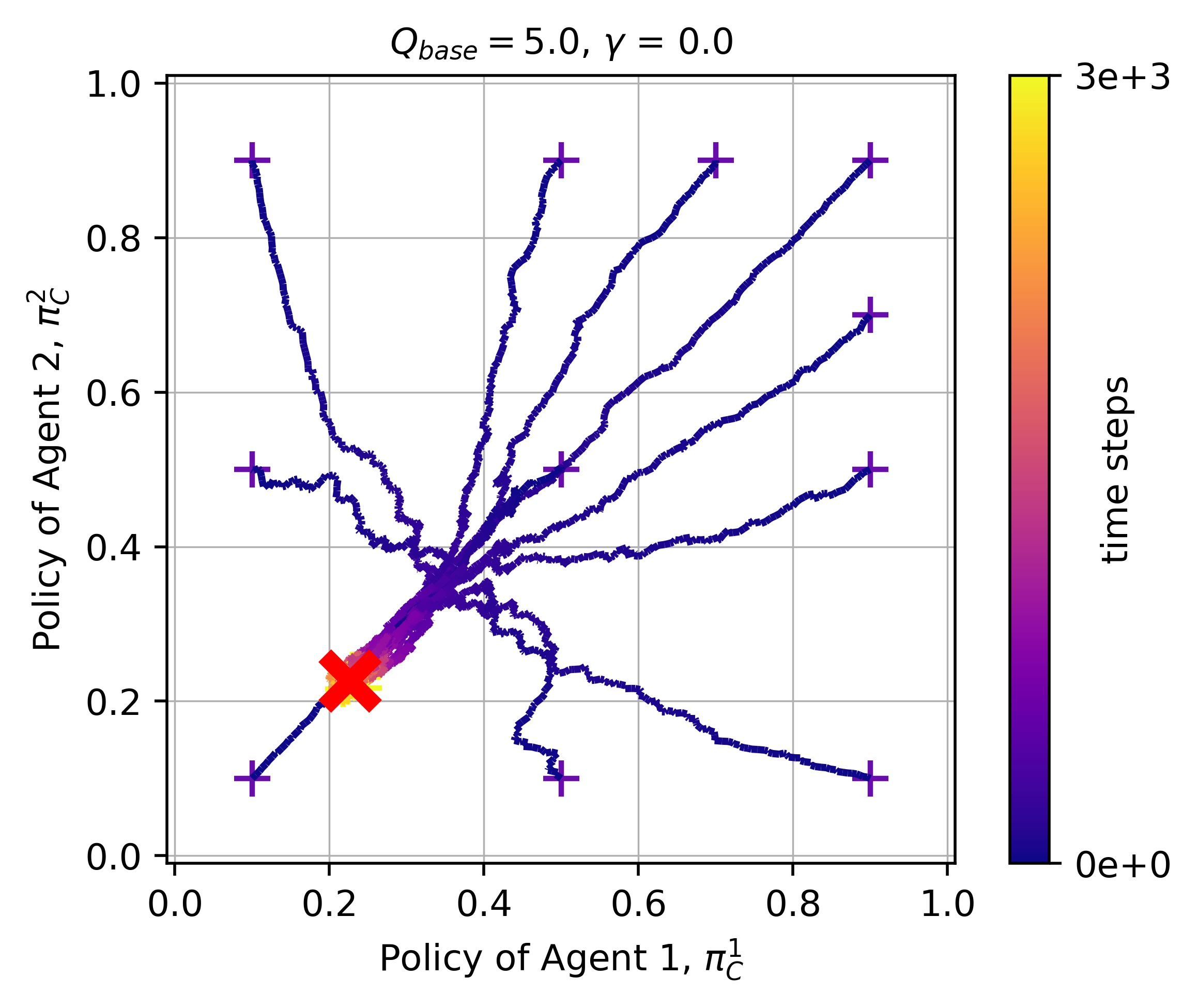}
        \put(\xdimI, \ydimI){C} % Top-left label
    \end{subfigure}
    \hfill
    \begin{subfigure}[b]{0.49\textwidth}
        \centering
        \includegraphics[width=\textwidth]{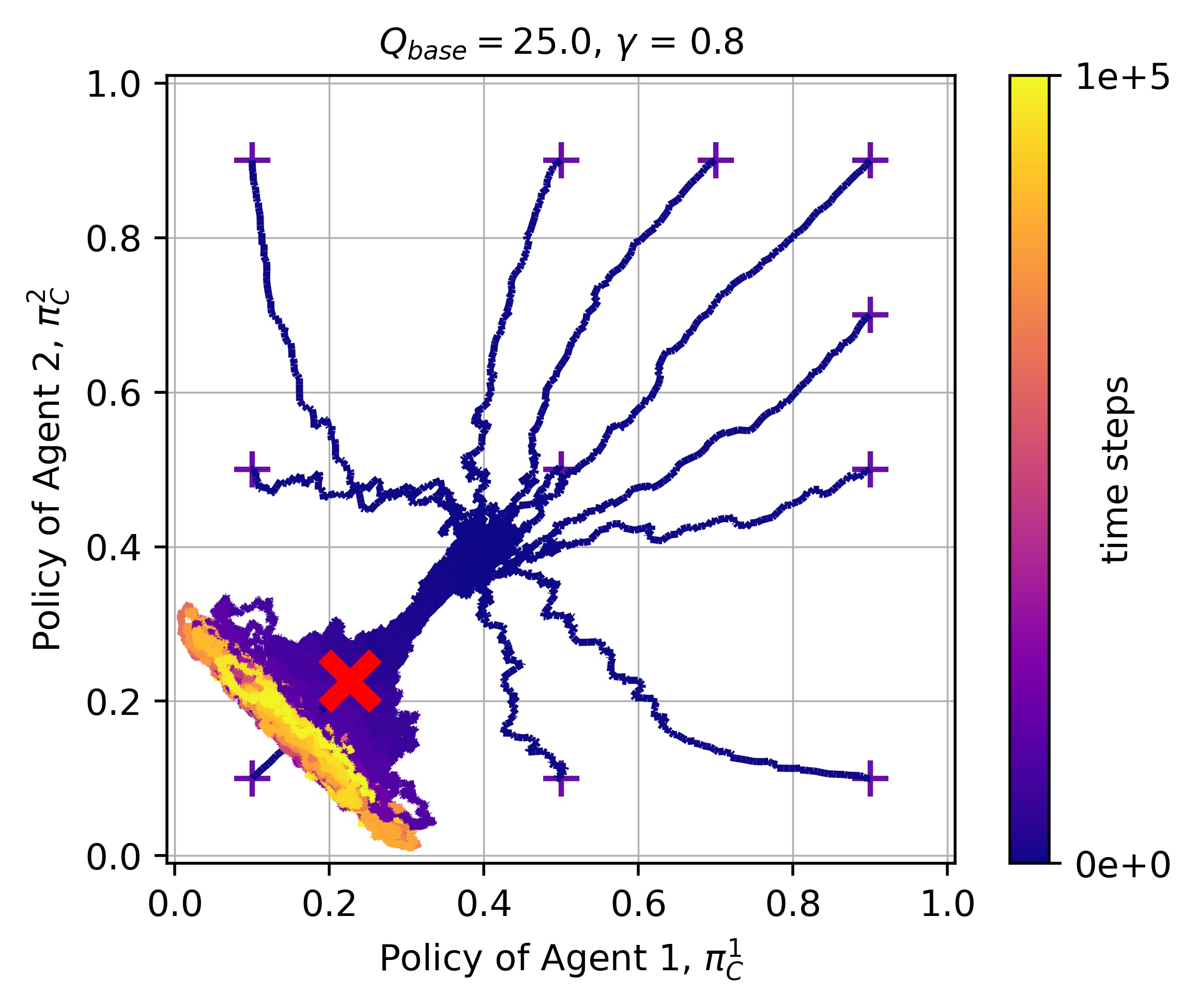}
        \put(\xdimI, \ydimI){D} % Top-left label
    \end{subfigure}
% MODELS
% Instructions: \xdim and \ydim should be set such that the labels are in the top left corner of each subfigure!
\newcommand{\xdimII}{-135}
\newcommand{\ydimII}{130}
    \centering
    \begin{subfigure}[t]{0.325\textwidth}
        \centering
        \includegraphics[width=\textwidth]{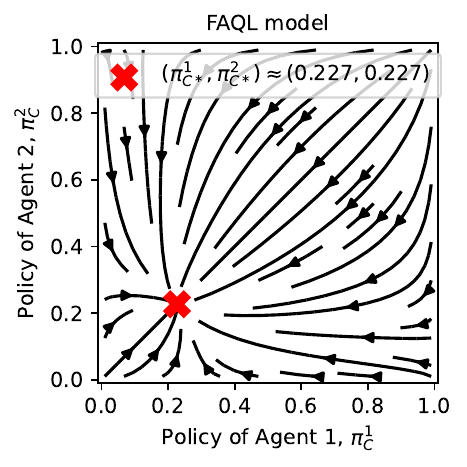}
        \put(\xdimII,145){\large \textbf{II}} % Top-left label for FIGURE II. Should be at the top left corner of the third row
        \put(\xdimII, \ydimII){E} % Top-left label
    \end{subfigure}
    \hfill
    \begin{subfigure}[t]{0.325\textwidth}
        \centering
        \includegraphics[width=\textwidth]{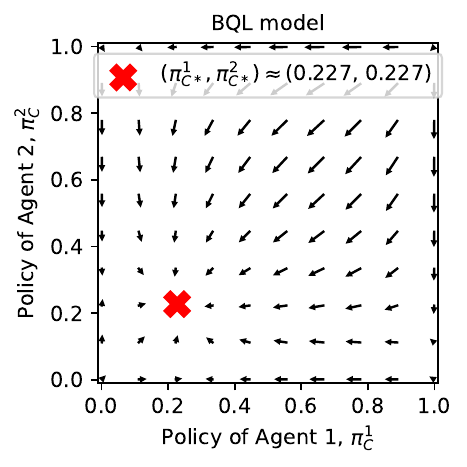}
        \put(\xdimII, \ydimII){F} % Top-left label
    \end{subfigure}
    \hfill
    \begin{subfigure}[t]{0.325\textwidth}
        \centering
        \includegraphics[width=\textwidth]{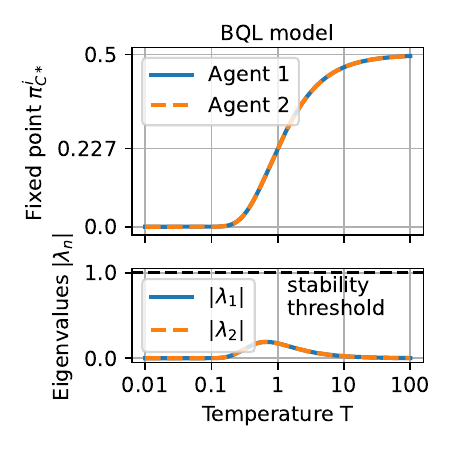}
        \put(\xdimII, \ydimII){G} % Top-left label
    \end{subfigure}
\caption{
Comparison between averaged policy trajectories of independent Q-learning on the Prisoner's Dilemma (\textbf{I}) and previous deterministic models (\textbf{II}) for $T=1$ and $\alpha=0.01$. 
\textbf{I:}
Top panels (A, B): $Q_{base} = \min(\mathbf{R}) / (1-\gamma)$.  
Bottom panels (C, D): $Q_{base} = \max(\mathbf{R}) / (1-\gamma)$.  
Left panels (A, C): $\gamma = 0$. Right panels (B, D): $\gamma = 0.8$. 
For each initialisation, five runs are executed. The trajectories from the same initialisation are grouped based on their final location in policy space (below or above the diagonal from (0,1) to (1,0)), and the mean of each group is plotted. Line thickness indicates the proportion of runs in each group. The colour gradient (purple to yellow) indicates time evolution. The red cross marks the fixed point of the FAQL/BQL model.
Note that for $Q_{base} = 0$ and $\gamma = 0.8$, some trajectories initialised in the top right appear to converge to the metastable phase of mutual cooperation in the depicted time span of $1 \times 10^5$ steps. 
\textbf{II:}
Vector fields of previous models. 
E: FAQL model in continuous time, defined by \eqref{eq:FAQL_model}. 
F: BQL model in discrete time, defined by \eqref{eq:BQL_model}. 
G: Stability analysis of the BQL model (see appendix \ref{sec:Appendix_BQL}). 
It has a unique symmetric fixed point $\boldsymbol{\pi}_* > 0$, depending on the temperature $T > 0$. 
All absolute eigenvalues of the Jacobian at $\pi^i_{C*}$ are below 1, indicating a stable node.
}
\label{fig:2}
\end{figure}
% -----------

% ----------- FIGURE 3 ------------
\begin{figure}[t]
% Instructions: \xdim and \ydim should be set such that the labels are in the top left corner of each subfigure!
\newcommand{\xdim}{-205}
\newcommand{\ydim}{170}
    \centering
    \begin{subfigure}[b]{0.49\textwidth}
        \centering
        \includegraphics[width=\textwidth]{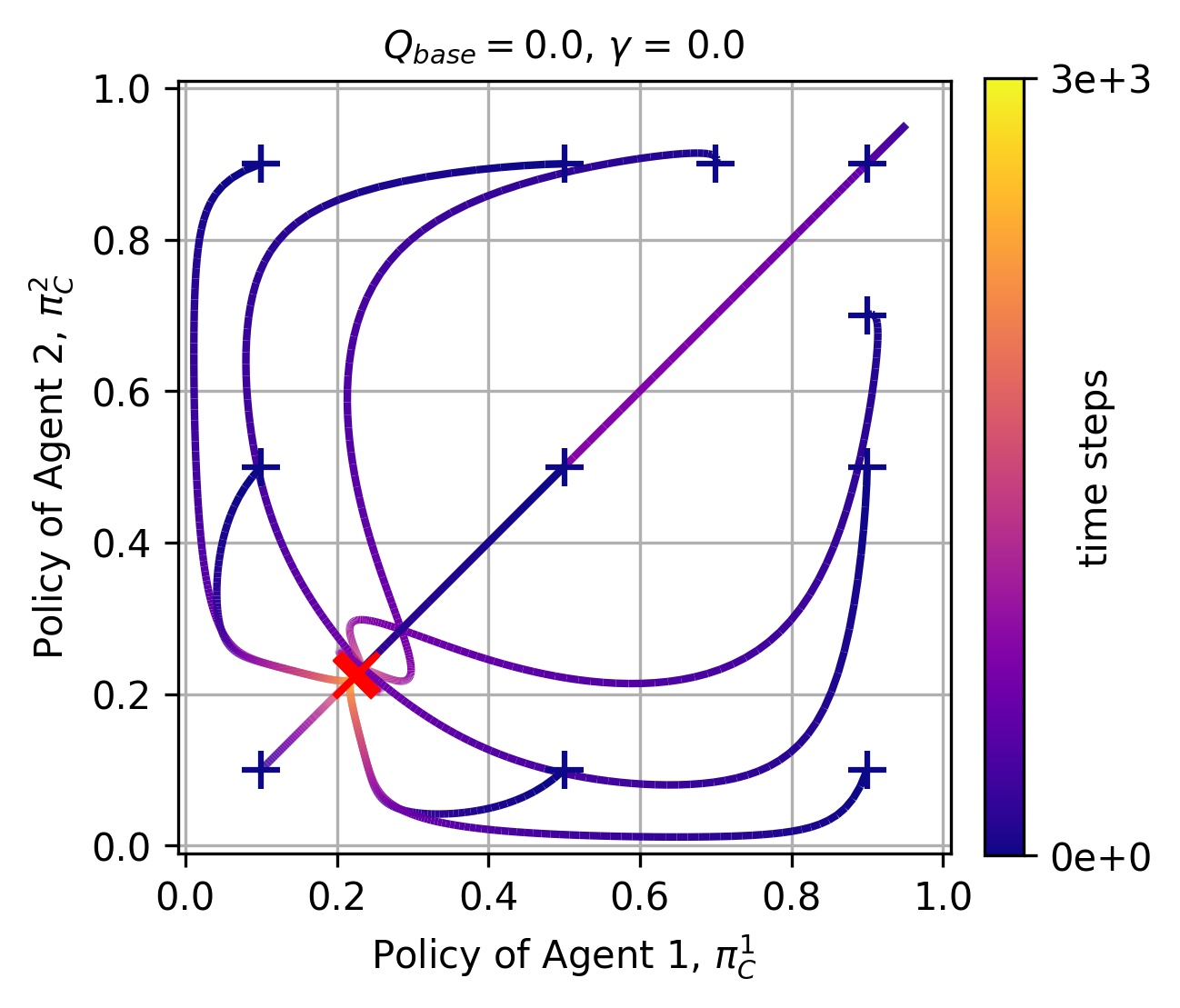}
        \put(\xdim, \ydim){A} % Top-left label
    \end{subfigure}
    \hfill
    \begin{subfigure}[b]{0.49\textwidth}
        \centering
        \includegraphics[width=\textwidth]{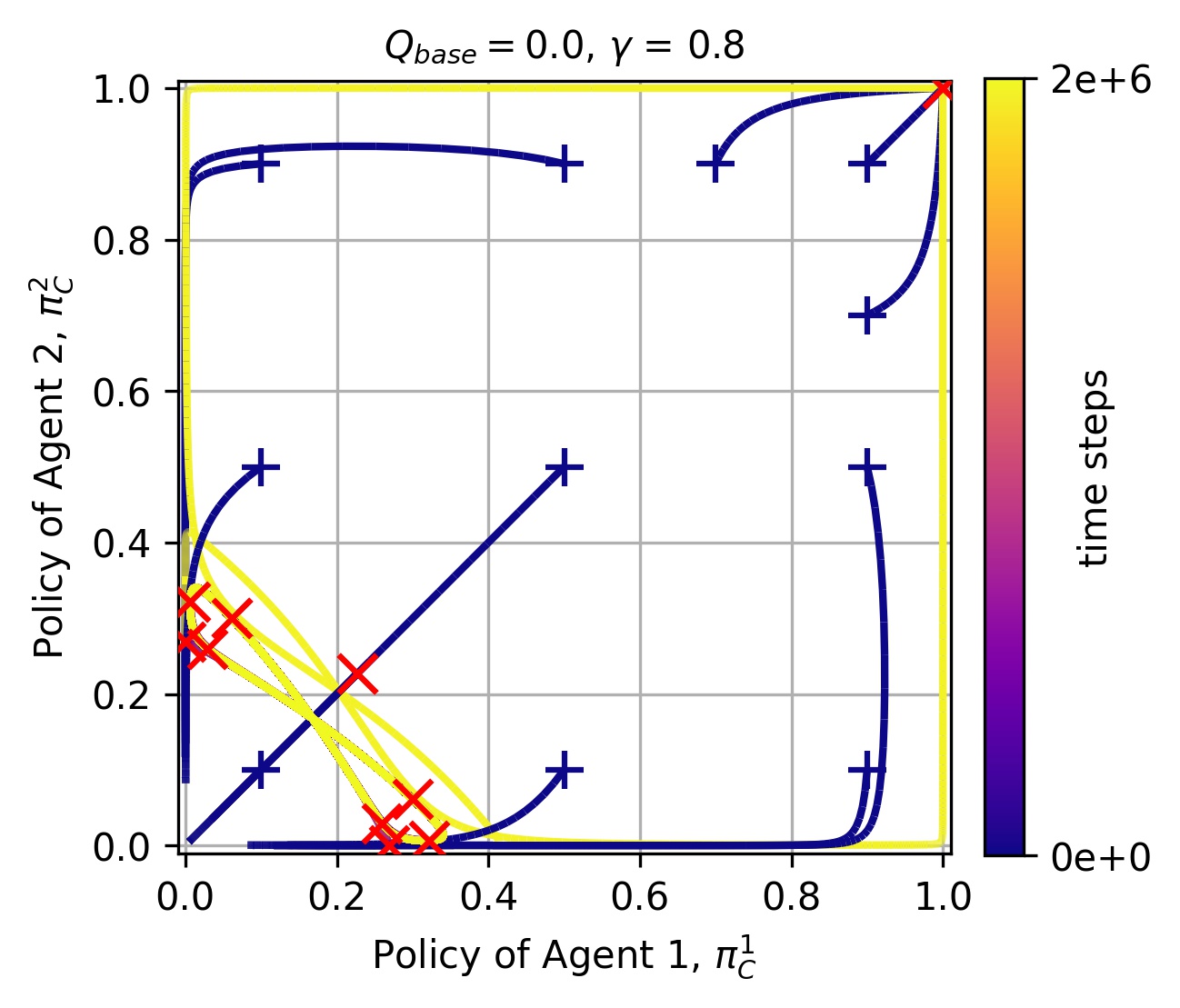}
        \put(\xdim, \ydim){B} % Top-left label
    \end{subfigure}
    \hfill
    \begin{subfigure}[b]{0.49\textwidth}
        \centering
        \includegraphics[width=\textwidth]{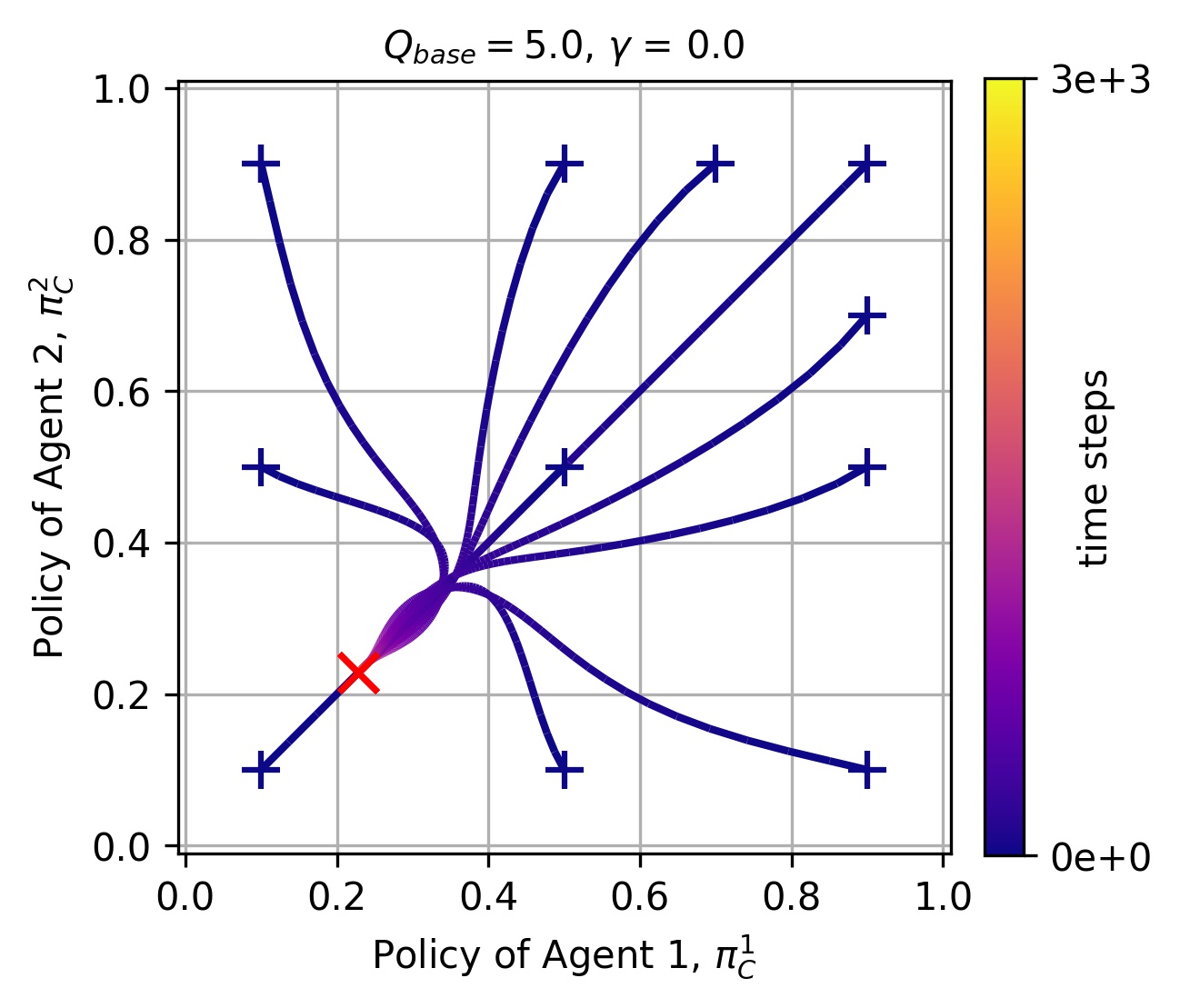}
        \put(\xdim, \ydim){C} % Top-left label
    \end{subfigure}
    \hfill
    \begin{subfigure}[b]{0.49\textwidth}
        \centering
        \includegraphics[width=\textwidth]{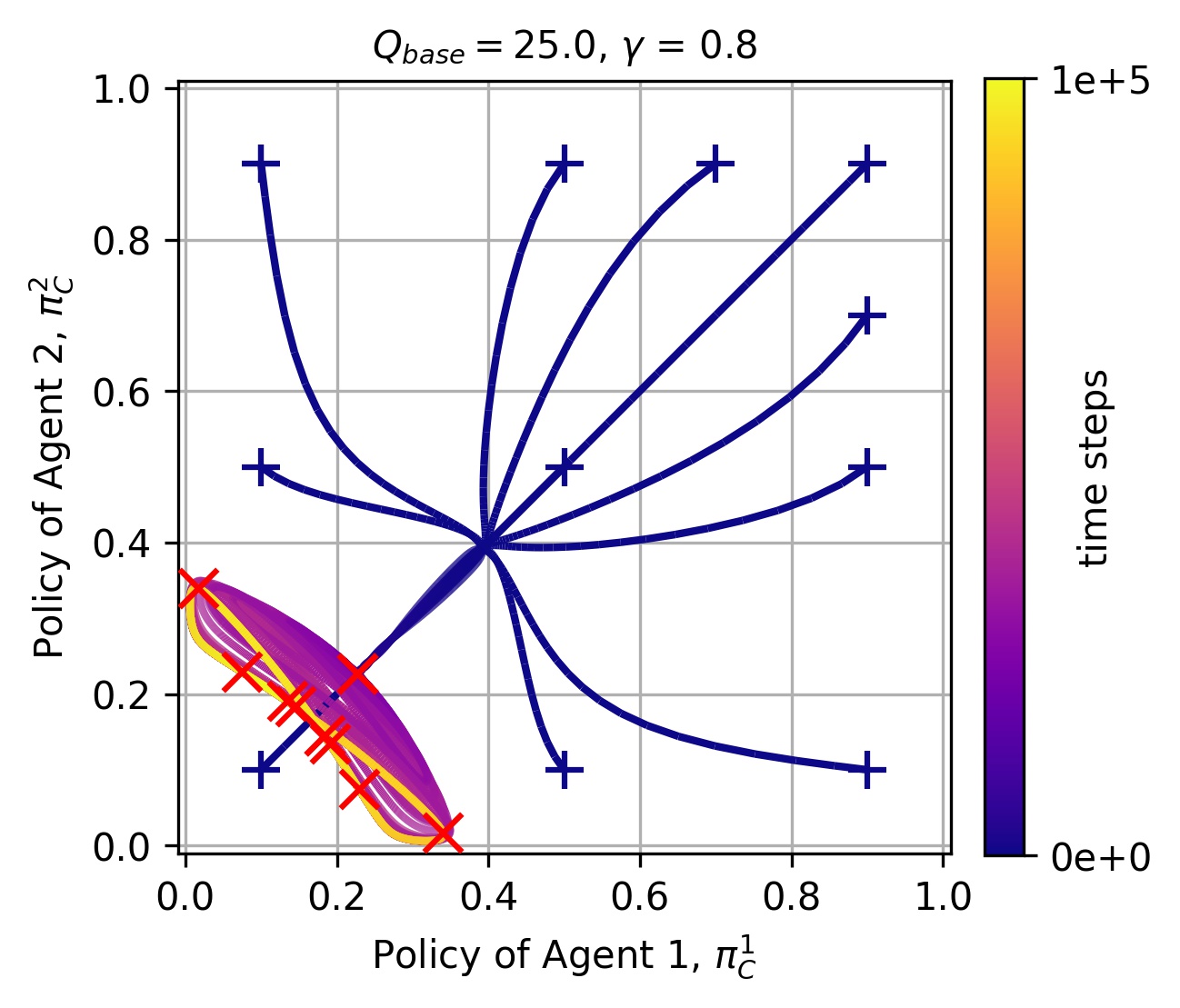}
        \put(\xdim, \ydim){D} % Top-left label
    \end{subfigure}
\caption{
Projection of our 4D deterministic approximation model of independent Q-learning on the Prisoner's Dilemma, defined by \eqref{eq:QLmodelNEW}, into 2D policy space for $T=1$, $\alpha=0.01$, and different values of $\gamma$ and $Q_{base}$.
The colour gradient (purple to yellow) represents time evolution. The end point of each trajectory is indicated by a red cross. 
Top panels (A, B): $Q_{base} = \min(\mathbf{R}) / (1-\gamma)$.  
Bottom panels (C, D): $Q_{base} = \max(\mathbf{R}) / (1-\gamma)$.  
Left panels (A, C): $\gamma = 0$. 
Right panels (B, D): $\gamma = 0.8$.
Note that in panel B, the trajectory initialised at $\pi^i_{C}(0) = 0.9$ eventually converges to the fixed point $\pi^i_{C*} \approx 0.227$, but only after $4 \times 10^7$ steps, far beyond the depicted $2 \times 10^6$ steps.
}
\label{fig:3}
\end{figure}
% -----------  

% -------------- Subsection --------------
\subsection{Comparison between the FAQL/BQL Model and Independent Q-learning}\label{sec:Discrepancies}
%- Mention how parameter values are handled
For convenience, we consider $\alpha^i, \gamma^i, T^i$ to be homogeneous. We set the learning rate to $\alpha = 0.01$ and the temperature to $T=1$, as we are mainly interested on the effect of the discount factor $\gamma$.

% - explain how initialisation is handled
To compare the algorithm, where learning occurs in Q-space, with the FAQL/BQL model, which describes learning in policy space, it is essential to understand how $Q$-values translate into policies and vice versa.
Note that an agent's probability to cooperate, $\pi^i_C$, does \emph{not} depend on the absolute $Q^i$-values but only on their difference, $ \Delta Q^i := Q^i_{D} - Q^i_{C}$, due to
\begin{equation}
\label{eq:BoltzmannPolicy_DeltaQ}
    \pi^i_C = \frac{e^{Q^i_C/T}}{e^{Q^i_C/T} + e^{ (Q^i_C + \Delta Q^i) / T}} = \frac{1}{1 + e^{ \Delta Q^i / T}}.
\end{equation}
Thus, a joint policy does \emph{not} correspond to a single point in Q-space but an affine subspace.
To study the influence of any initial policy $\boldsymbol \pi(0)$ on the algorithmic learning process, we first need to specify initial $\mathbf Q(0)$-values which fulfil $\boldsymbol \pi(0) = f(\mathbf Q(0))$. 
To this end, we define for any given $\pi^i(0)$
\begin{equation}
\label{eq:Q-value_initialisation}
    \begin{aligned}
    Q^i_{C}(0) &:= Q_{base} - \frac{\Delta Q^i (\pi^i_C(0)) }{2} ,\\
    Q^i_{D}(0) &:= Q_{base} + \frac{\Delta Q^i (\pi^i_C(0)) }{2} 
    ,
\end{aligned}
\end{equation}
where $Q_{base}$ is a parameter that governs the overall initial level of $Q$-values. 

% - Detailed Description of Figure 1
Figure \ref{fig:1}.A and \ref{fig:1}.B depict the time evolution of single runs of independent Q-learning for $Q_{base} = 0$, $\gamma = 0.8$ and two different initial joint policies.
In both cases, after the first few hundred time steps, the policy trajectories settle into metastable phases where they remain for an extended period. After a very long time, the behaviour undergoes a drastic shift, and the policies transition into a sustained oscillatory pattern that persists indefinitely.  
For the initial joint policy $(\pi^1_C(0), \pi^2_C(0)) = (0.5, 0.48)$, this transition occurs after approximately $70$ thousand steps. For $(\pi^1_C(0), \pi^2_C(0)) = (0.9, 0.7)$, the shift is even more pronounced. Initially, the policies seem to converge on mutual cooperation, which appears to contradict individually rational behaviour. However, after about \emph{two million} steps, the trajectories descend into the same asymmetric metastable phase observed for $(\pi^1_C(0), \pi^2_C(0)) = (0.5, 0.48)$, before ultimately transitioning into the indefinite oscillations.
In stark contrast, the FAQL/BQL models predict fundamentally simpler behaviour: as shown by the dotted lines in Figure \ref{fig:1}.C and \ref{fig:1}.D, their policy trajectories quickly converge to a joint policy within just a few hundred steps. That the previous models do not describe actual Q-learning can also be seen in figure \ref{fig:2}. 

Figure \ref{fig:2}.II displays the dynamics of the simplified models. A linear stability analysis of the BQL model (see appendix \ref{sec:Appendix_BQL}) shows that the system has a unique stable fixed point, ranging from $\lim_{T \rightarrow \infty} \pi^i_{C*} = 0.5$ to $\lim_{T \rightarrow 0} \pi^i_{C*} = 0$, which is the Nash equilibrium: both agents choose to defect. 

Figure \ref{fig:2}.I shows averaged policy trajectories of Q-learning over five runs for two different initialisation approaches and two different values of $\gamma$.
For $ Q_{base} = \min(\mathbf{R}) / (1-\gamma) = 0 $, the trajectories deviate from the model, following the edges of the policy space instead. 
For $ Q_{base} = \max(\mathbf{R}) / (1-\gamma) $, the trajectories initially cluster near the center of the policy space.
For $\gamma = 0$, although the trajectories differ from the FAQL/BQL model, they eventually equilibrate around the fixed point, regardless of initialisation. However, for $\gamma = 0.8$ the trajectories fall into indefinite oscillations, which are not centred around the fixed point.
For $Q_{base} = 0, \gamma = 0.8$, some trajectories appear to converge to mutual cooperation in the depicted time span of $1 \times 10^5$ steps. However, as mentioned above, these states are only metastable. Given sufficient time, the trajectories eventually transition to the same oscillatory pattern observed in other trajectories. Notably, these metastable phases do \emph{not} occur for trajectories initialised at $ Q_{base} = 25 $. \\
\\
% - KEY TAKE AWAYS
In summary, the stylised discrepancies are:
\begin{enumerate}
    \item 
    Whereas the FAQL/BQL model dynamics converge to a single Logit Quantal Response equilibrium in the Prisoner's Dilemma after a couple of hundred steps, actual independent Q-learning does \emph{not} necessarily converge to any strategic equilibrium and may instead settle into oscillations that emerge only after millions of steps. 
    
    \item 
    Whereas the FAQL/BQL model reside in the lower-dimensional policy space, actual independent Q-learning dynamics \emph{cannot} be reduced from the higher-dimensional Q-space: the initialisation ($Q_{base}$) matters.
    
    \item 
    Whereas the FAQL/BQL model is independent of the discount factor in single-state environments, actual independent Q-learning dynamics are clearly influenced by changes in $\gamma$ and exhibit fundamentally different behaviour.

\end{enumerate}

\section{A Choice-Probability-Aware Model of Independent Q-learning} \label{sec:ourModel}

% - Einordnung in bisherige Literatur
The discrepancies between the FAQL/BQL model and independent Q-learning arise from the implicit assumption that \emph{all} $Q$-values are updated at each step. Some researchers recognised the need to consider update frequencies but modified the algorithm to fit the model, rather than adjusting the model itself \cite{leslie2005individual, kaisers2010frequency, Barfuss2019}. 
%While some publications \cite{leslie2005individual, kaisers2010frequency, bloembergen2015evolutionary, hernandez2017survey, Barfuss21} acknowledged that the FAQL model \cite{tuyls_2003} does not accurately represent actual independent Q-learning, others---including \cite{Kianercy_2012}, \cite{galstyan2013continuous}, and more recent works like \cite{leonardos2022exploration} and \cite{mintz2024evolutionary}---do not mention this discrepancy, potentially overlooking its implications.
Recently in \citeyear{hu2022dynamics}, \citeauthor{hu2022dynamics} proposed an adjusted ``continuity equation model" of independent Q-learning in large-scale multi-agent systems modelled as population games \citep{hu2022dynamics}.  
However, their model is limited to the case $\gamma = 0$. Thus, we cannot apply it to explain all of the stylised discrepancies from above. 

Here, we propose an approximation model for independent Q-learning in a single-state, repeated environment, with discounting but no memory, as defined in section \ref{sec:IndependentQLearning}. 
We show that all stylised discrepancies between actual independent Q-learning and previous approximation models can be explained by adjusting the previous models' update frequencies to be proportional to the current agent's policies.
Our primary focus is then to demonstrate and rigorously prove that heterogeneous update frequencies can fundamentally alter a system's behaviour, emphasising the need for caution when using MARL as a modelling tool.

%- Nessecary Assumption: Small learning rate! (like in FAQ model)
% - Explanation why discrete and not continuous time
We construct our deterministic approximation by isolating the dynamics between consecutive time steps. At each step, we study the expectation of the next step given the current values. The model stays in discrete-time, aligning closer with the inherent nature of computer simulations. 
Importantly, this approach replaces the Kronecker delta $ \delta_{A^i(t) a^i} $ in \eqref{eq:StochasticQ-UpdateRule} with the probability---or update frequency---$ \pi^i_{a^i}(t) $, leading to
\begin{equation}
\label{eq:QLmodelNEW}
    \begin{aligned}
        \mathbb E_{ \mathbf A(t) \sim \boldsymbol \pi(t) } [ Q^i_{a^i}(t+1) \mid Q^i(t) ] &= 
        Q^i_{a^i}(t) \\
        &\quad
        + \alpha
        \pi^i_{a^i}(t) 
        \biggl [
        \mathbb E_{ A^{-i}(t) \sim \pi^{-i}(t) }
        R^i_{a^i A^{-i}(t)}
        + \gamma \max_{b^i \in \mathcal A^i} Q^i_{b^i}(t) 
        - Q^i_{a^i}(t) 
        \biggr ]
        .
    \end{aligned}
\end{equation} 
It is \emph{not possible} to reduce these dynamics into policy space, as done in the FAQL/BQL model. When attempting to transform \eqref{eq:QLmodelNEW} into $\Delta Q$-space for the Prisoner's Dilemma,
\begin{equation*}
\begin{aligned}
\label{eq:QLmodelDELTAQ}
    \mathbb{E}_{\mathbf A(t) \sim \boldsymbol \pi^{-i}(t)} 
    \Delta Q^i ( t+1 ) 
    &=  
    \mathbb{E}_{\mathbf A(t) \sim \boldsymbol \pi^{-i}(t)} 
    Q^i_{D}(t+1) 
    - \mathbb{E}_{\mathbf A(t) \sim \boldsymbol \pi^{-i}(t)} 
    Q^i_{C}(t+1)
    \\
    &=
    \Delta Q^i ( t ) 
    + \alpha 
    \biggl [ 
    \mathbb{E}_{A^{-i}(t) \sim \pi^{-i}(t)}
    \biggl (
    \pi^i_{D}(t)
    R^i_{a^i=D, A^{-i}(t)} 
    - 
    \pi^i_{C}(t)
    R^i_{a^i=C,  A^{-i}(t)} 
    \biggr )
    \\
    & \quad
    +
    \left ( \pi^i_{D}(t) - \pi^i_{C}(t) \right )
    \gamma \max_{b^i \in \mathcal A^i} Q^i_{b^i}(t) 
      -
    \pi^i_{D}(t)
    Q^i_{D} (t) 
    +
    \pi^i_{C}(t)
    Q^i_{C} (t) 
    \biggr ] 
    ,
\end{aligned}
\end{equation*}
it becomes apparent, that the last three terms cannot be expressed in terms of $\Delta Q^i$ because of the update frequencies $\pi^i_{a^i}(t) \neq 1$. 

However a \emph{projection} of the deterministic dynamics into policy space can still be illustrated (figure \ref{fig:3}). 
For reasonably small learning rates ($\alpha = 0.01$), a comparison with the averaged trajectories of Q-learning (figure \ref{fig:2}) demonstrates that \eqref{eq:QLmodelNEW} captures the observed complexities. For $\gamma = 0$, all trajectories converge to the fixed point of the FAQL/BQL model, $\pi^i_{C*} \approx 0.227$. In contrast, for $\gamma = 0.8$, the behaviour depends on the initial policies: symmetric initial policies converge to $\pi^i_{C*}$ while asymmetric initial policies lead to oscillatory dynamics.
Note that for $Q_{base} = 0, \gamma = 0.8$, the trajectory starting at the symmetric initial condition $\pi^i_C = 0.9$ remains in the cooperation state for up to two million steps, seemingly contradicting the statement just made. However, after an astonishing four \emph{billion} steps, it finally converges to $\pi^i_{C*}$. 
These phenomena are readily explained and proven through a stability analysis of our model, offering an efficient approach while avoiding ambiguities of interpreting individual trajectories or specific parameter cases.

% ------------------------- FIGURE 4 ----------------------------- 
\begin{figure}[t]
% Instructions: \xdim and \ydim should be set such that the labels are in the top left corner of each subfigure!
\newcommand{\ydim}{210}
\centering
\includegraphics[width=\textwidth]{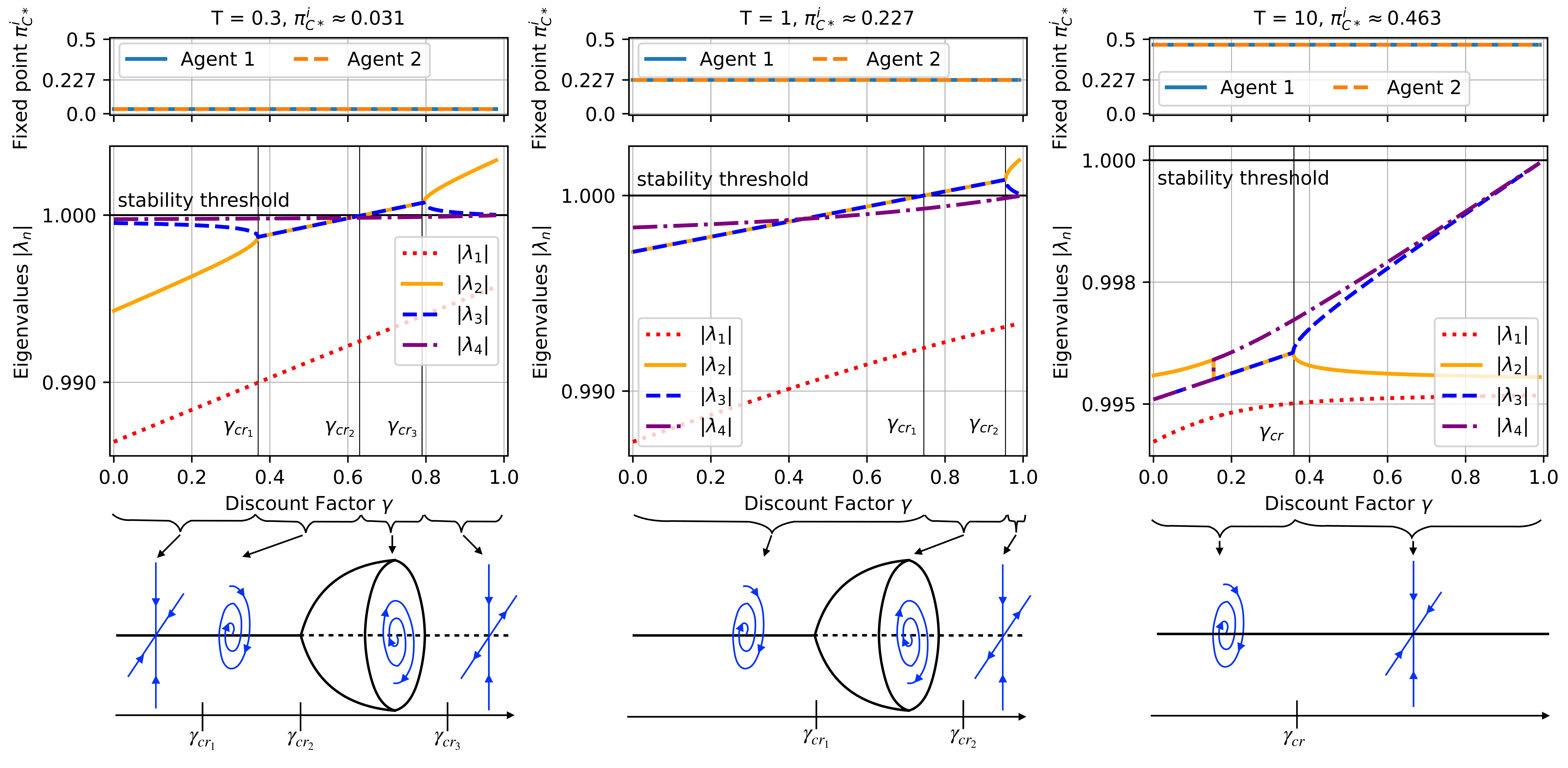}
\put(-410, \ydim){A} % Top-left label
\put(-270, \ydim){B} % Top-left label
\put(-130, \ydim){C} % Top-left label
\caption{Stability analysis of our model, defined by \eqref{eq:QLmodelNEW}, for $\alpha = 0.01$ and three different temperature values: $T=0.3$ (A), $T=1$ (B), and $T=10$ (C).  
The deterministic 4D system shares the same unique symmetric fixed point $\boldsymbol{\pi}_* = \boldsymbol{\pi}(\mathbf {Q}_*)$ in \emph{policy space} as the 2D FAQL/BQL model (figure \ref{fig:2}).  
The first row shows the position of the 4D fixed point $\bf Q_*$, defined by \eqref{eq:FixedPointQ}, in 2D policy space.
Specifically, it illustrates how the projected equilibrium policy $\pi^i_{C*} := \pi^i_{C}(\mathbf {Q}_*)$ is not affected by the discount factor. 
The second row shows the absolute eigenvalues of the Jacobian matrix at the 4D fixed point $\mathbf {Q}_*$ as a function of $\gamma$, with the stability threshold ($|\lambda| = 1$) highlighted. It demonstrates that although the position of the fixed point in policy space remains unaffected by $\gamma$, its stability properties changes.
For instance, at $T=1$, the dynamics undergoes a supercritical Neimark-Sacker bifurcation at $\gamma_{cr_1} \approx 0.75$.
The third row provides schematic representations of the corresponding dynamical regimes for different ranges of $\gamma$, illustrating transitions between stability, oscillatory dynamics, and divergence.  
}
\label{fig:4}
\end{figure}

% -------------------------- FIGURE 5 ----------------------------- 
\begin{figure}
% Instructions: \xdim and \ydim should be set such that the labels are in the top left corner of each subfigure!
\newcommand{\xdimI}{-130}
\newcommand{\xdimII}{-130}
\newcommand{\xdimIII}{-130}
\newcommand{\ydimI}{135}
\newcommand{\ydimII}{140}
\newcommand{\ydimIII}{100}
    \begin{minipage}[t]{0.32\linewidth}  % Adjust the width of each minipage as needed
        \centering
        \begin{subfigure}[b]{\textwidth}
            \centering
            \includegraphics[width=\textwidth]{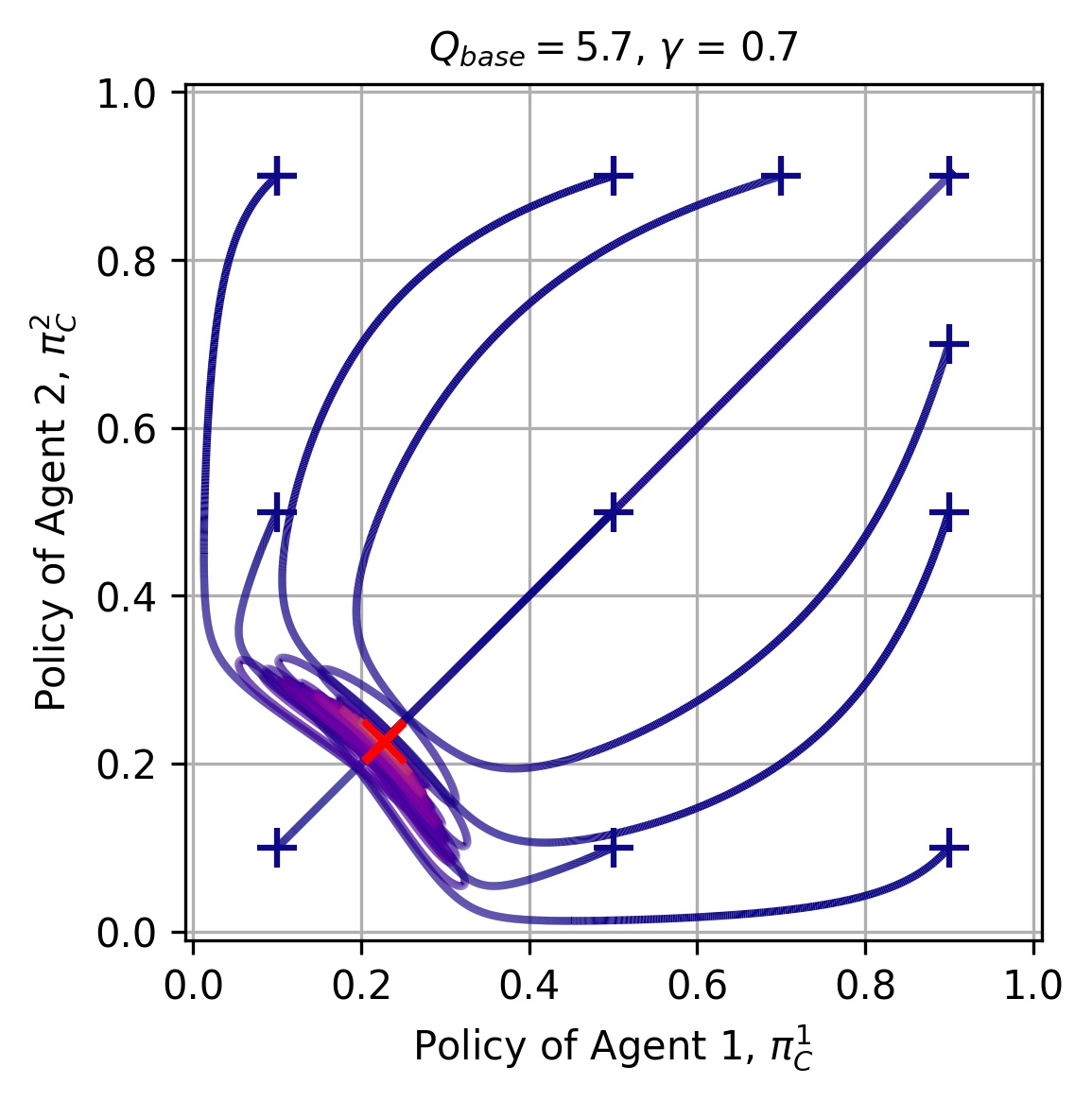}
            \put(\xdimI, \ydimI){A} % Top-left label
        \end{subfigure}
        \hfill
        \begin{subfigure}[b]{\textwidth}
            \centering
            \includegraphics[width=\textwidth]{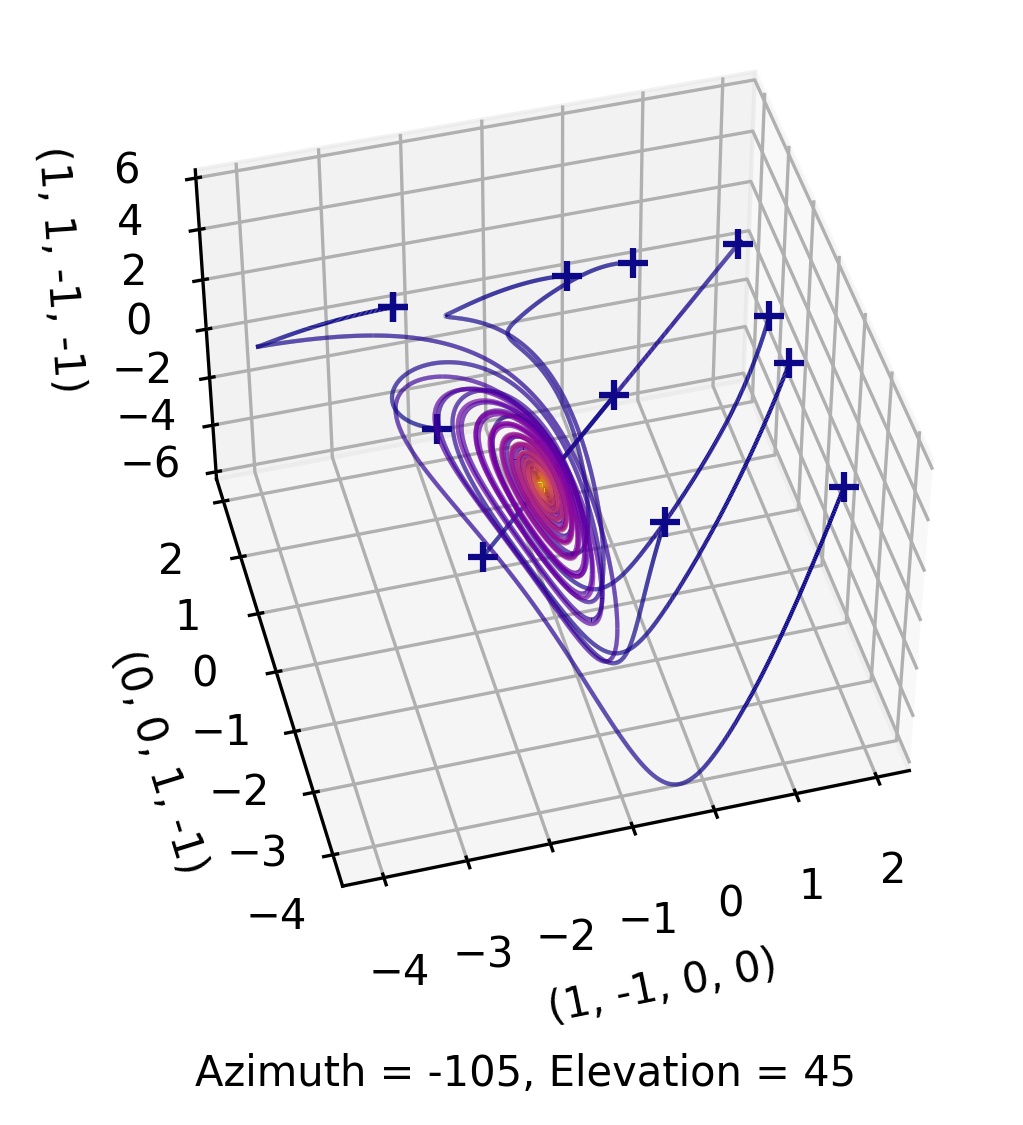}
            \put(\xdimI, \ydimII){D} % Top-left label
        \end{subfigure}
        \hfill
        \begin{subfigure}[b]{\textwidth}
            \centering
            \includegraphics[width=\textwidth]{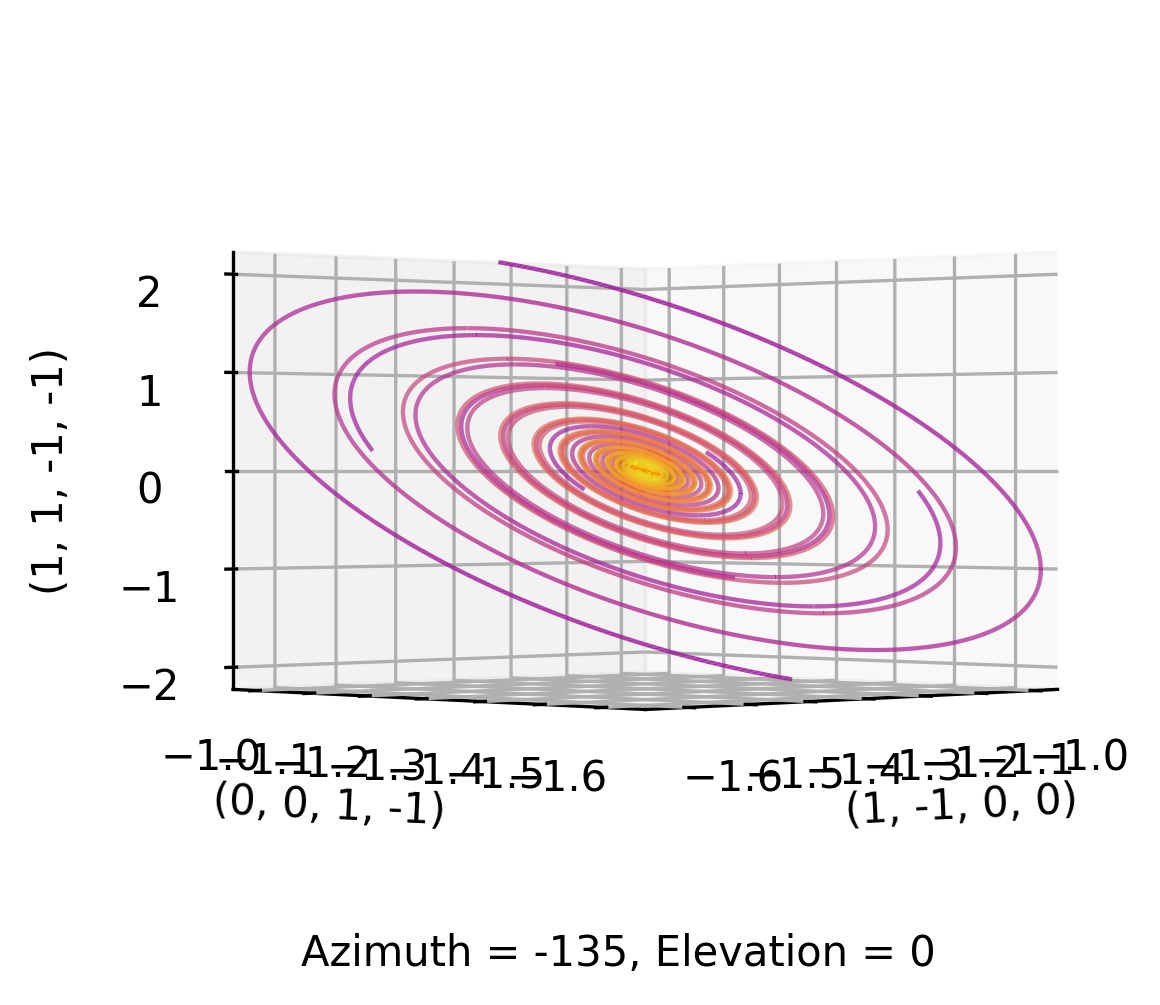}
            \put(\xdimI, \ydimIII){G} % Top-left label
        \end{subfigure}
    \end{minipage}
    \begin{minipage}[t]{0.32\linewidth}  % Adjust the width of each minipage as needed
        \centering
        \begin{subfigure}[b]{\textwidth}
            \centering
            \includegraphics[width=\textwidth]{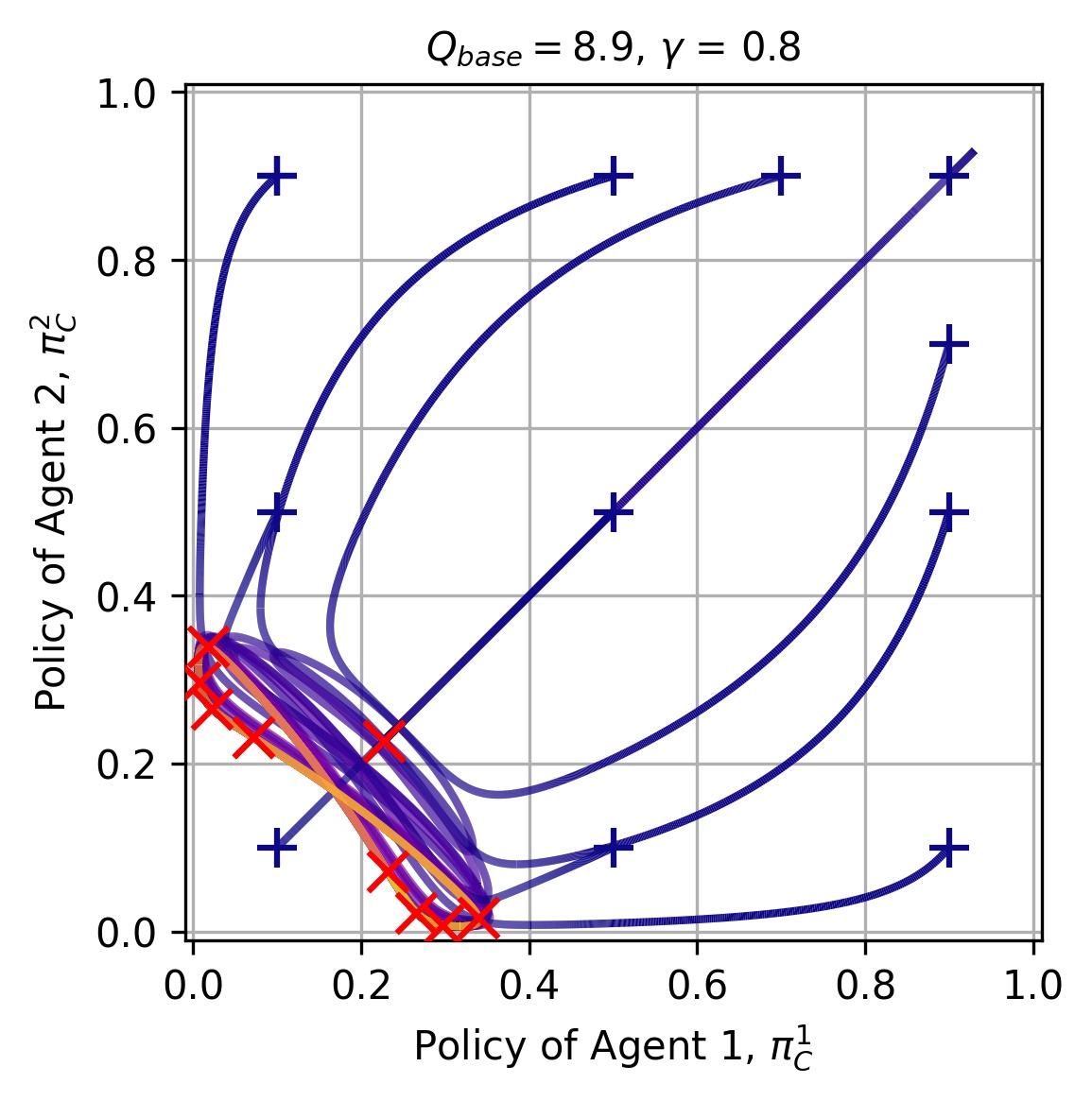}
            \put(\xdimII, \ydimI){B} % Top-left label
        \end{subfigure}
        \hfill
        \begin{subfigure}[b]{\textwidth}
            \centering
            \includegraphics[width=\textwidth]{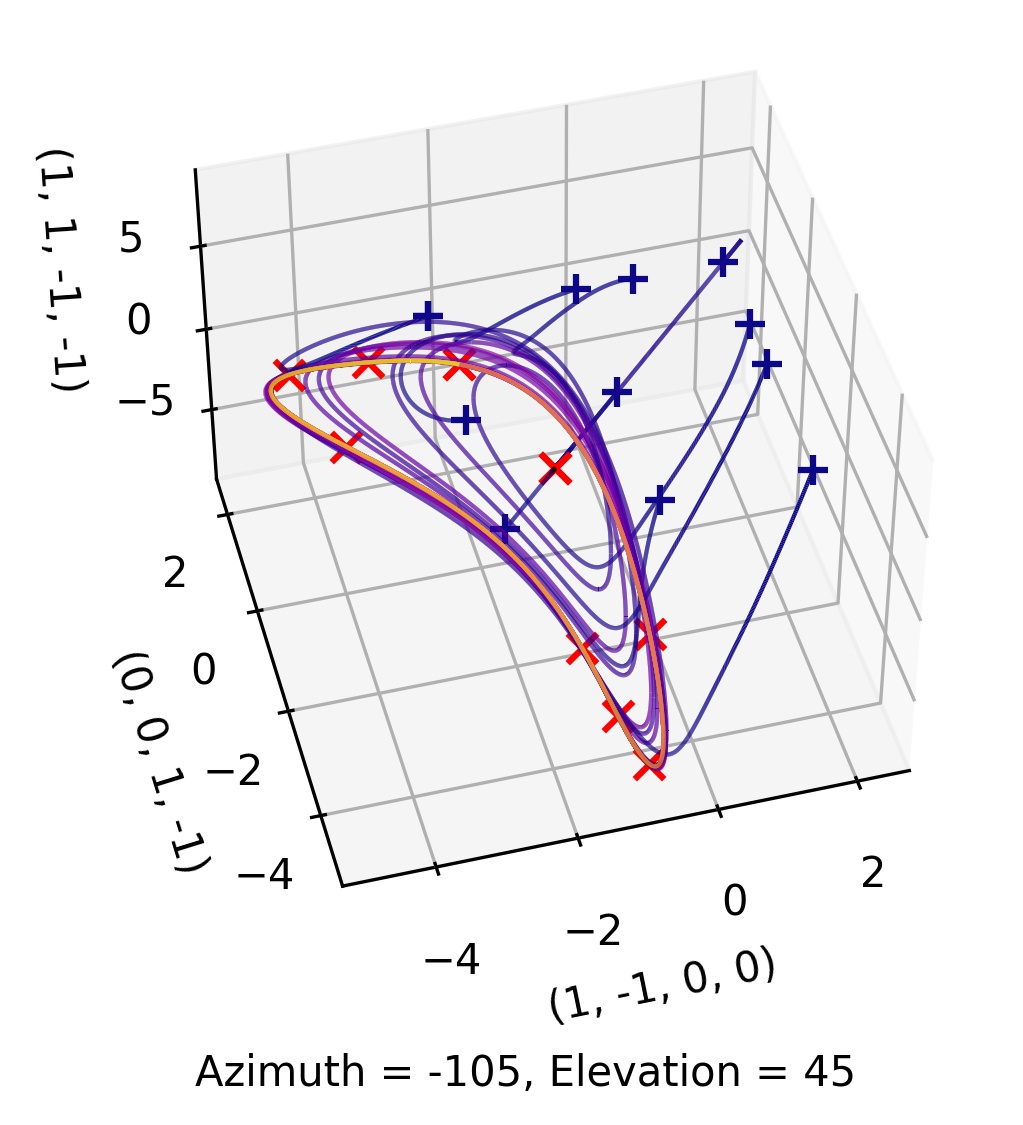}
            \put(\xdimII, \ydimII){E} % Top-left label
        \end{subfigure}
        \hfill
        \begin{subfigure}[b]{\textwidth}
            \centering
            \includegraphics[width=\textwidth]{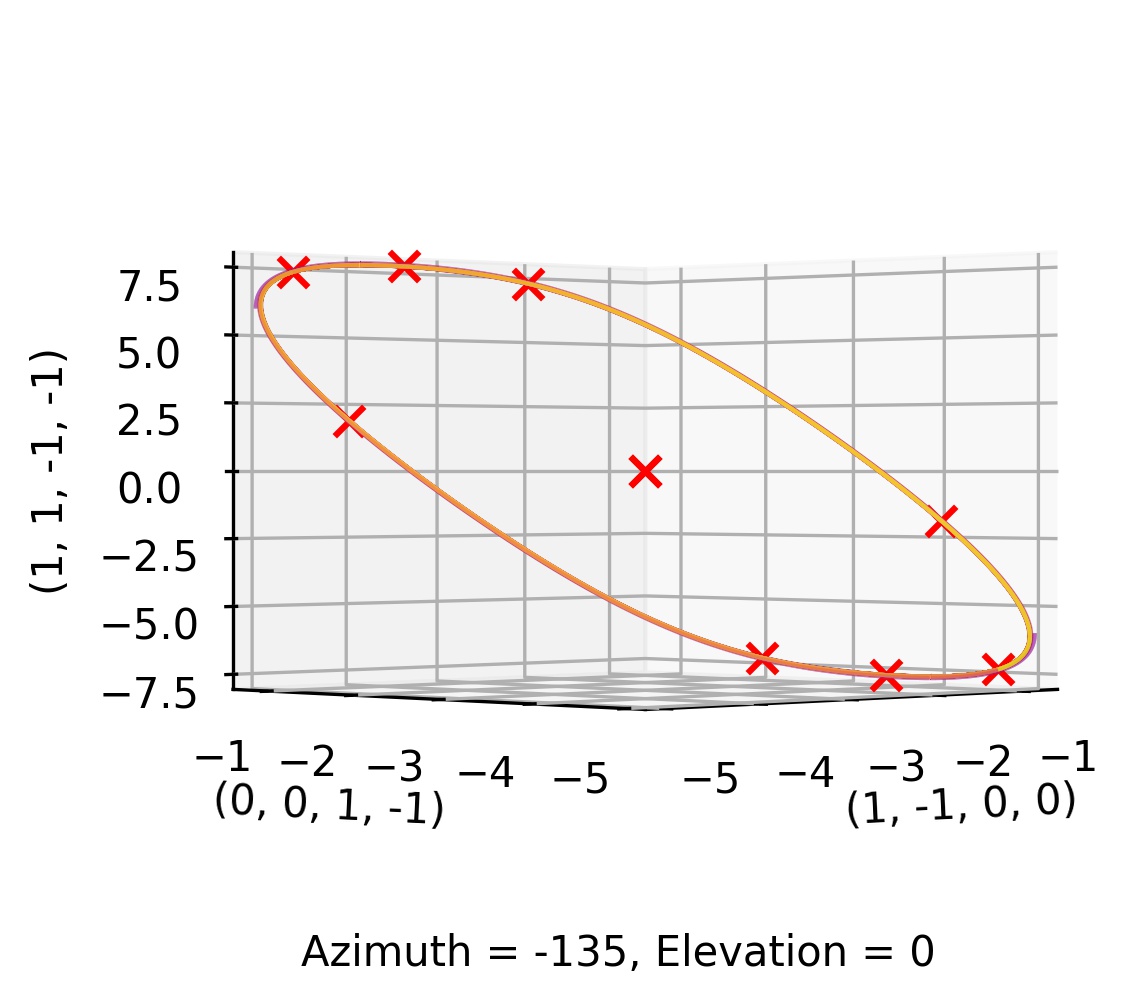}
            \put(\xdimII, \ydimIII){H} % Top-left label
        \end{subfigure}
    \end{minipage}
    \begin{minipage}[t]{0.32\linewidth}  % Adjust the width of each minipage as needed
        \centering
        \begin{subfigure}[b]{\textwidth}
            \centering
            \includegraphics[width=\textwidth]{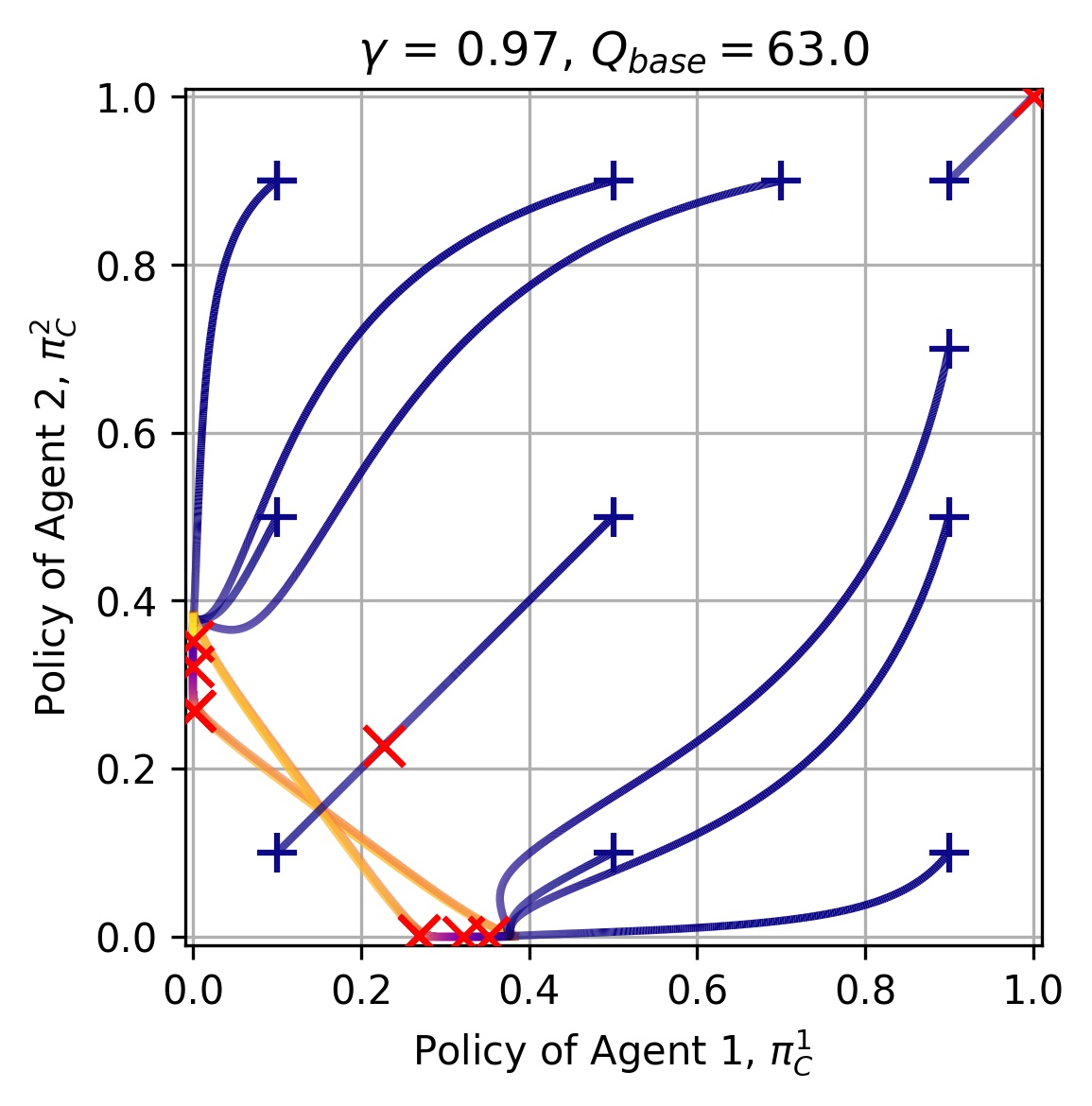}
            \put(\xdimIII, \ydimI){C} % Top-left label
        \end{subfigure}
        \hfill
        \begin{subfigure}[b]{\textwidth}
            \centering
            \includegraphics[width=\textwidth]{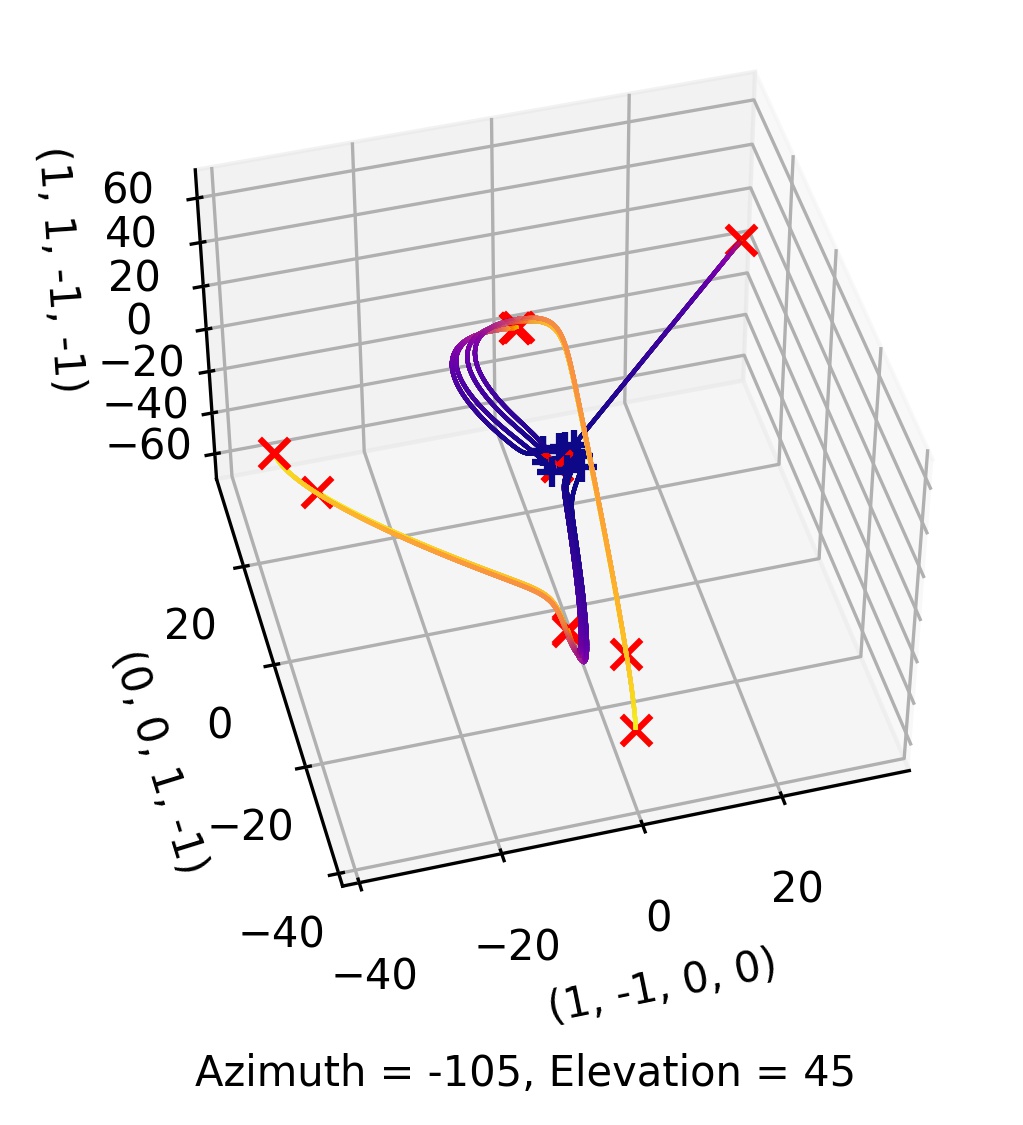}
            \put(\xdimIII, \ydimII){F} % Top-left label
        \end{subfigure}
        \hfill
        \begin{subfigure}[b]{\textwidth}
            \centering
            \includegraphics[width=\textwidth]{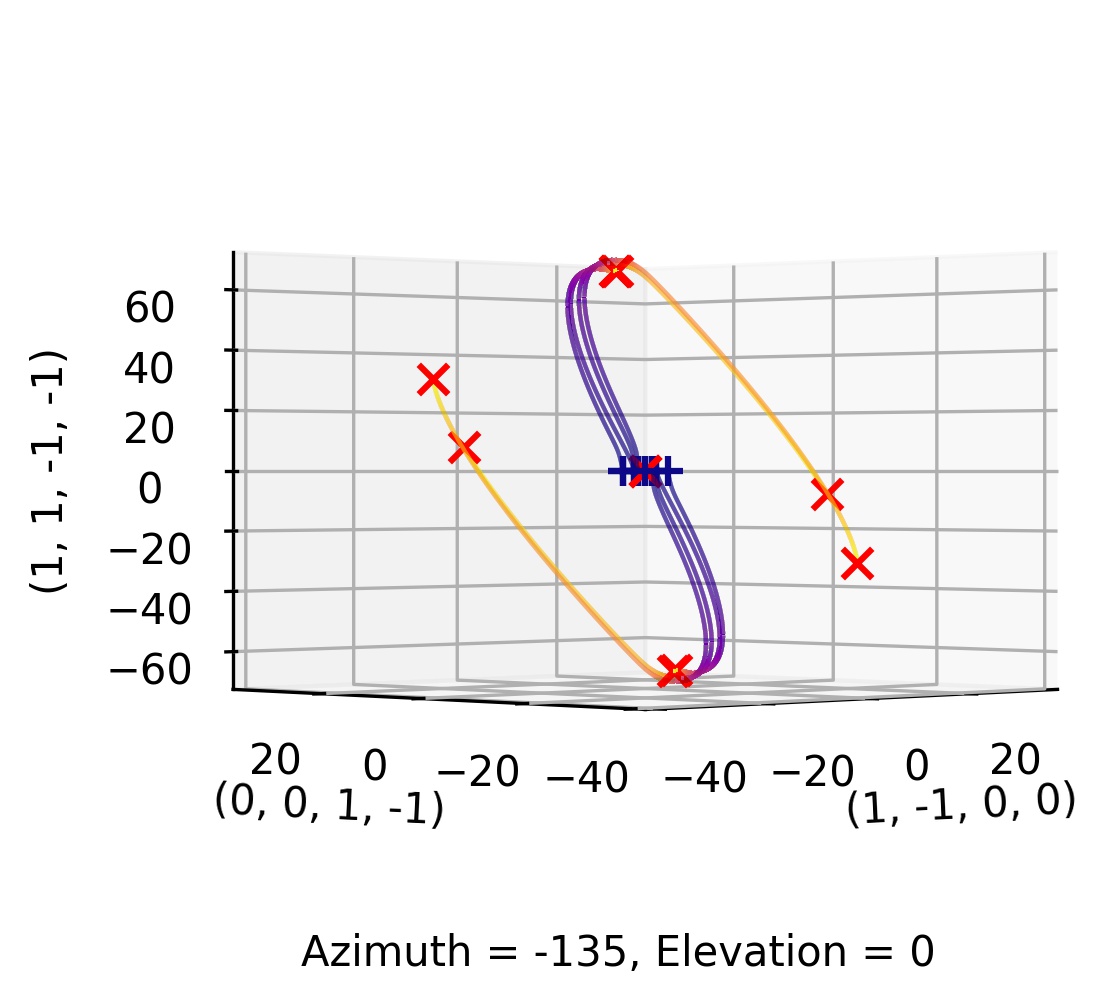}
            \put(\xdimIII, \ydimIII){I} % Top-left label
        \end{subfigure}
    \end{minipage}
\caption{Projection of 4D deterministic dynamics of independent Q-learning on the Prisoner's Dilemma, defined by \eqref{eq:QLmodelNEW}, for $T=1$, $\alpha=0.01$ and different values of $\gamma$. 
Left panels (A, D, G): $\gamma = 0.7$.
Middle panels (B, E, F): $\gamma = 0.8$.
Right panels (G, H, I): $\gamma = 0.97$.
All trajectories are initialised around the fixed point $Q$-values, defined by \eqref{eq:FixedPointQ}: $Q_{base} = Q_{C*} + \Delta Q_*/2$.
The colour gradient (purple to yellow) represents time evolution over $3 \times 10^4$ steps. The end point of each trajectory is indicated by a red cross.
Top panels (A, B, C): Projection of 4D dynamics into 2D policy space.
Middle panels (D, E, F): Projection into a 3D space defined by the basis vectors $\mathbf{q}_1 = (1, -1, 0, 0)$, $\mathbf{q}_2 = (0, 0, 1, -1)$, and $\mathbf{q}_3 = (1, 1, -1, -1)$.  
The first two dimensions represent the $\Delta Q^i$-values, while the third dimension captures the difference between agents.
Bottom panels (G, H, I): Projection into the same 3D space, viewed from a different angle.
For $\gamma = 0.7$ and $\gamma = 0.8$, only the last two-thirds of the time evolution are shown for clarity.
For $\gamma = 0.7$, the unique fixed point $\pi^i_{C*}$ is a stable focus. For $\gamma = 0.8$, it is an unstable focus surrounded by a stable limit cycle for all asymmetric joint policies. For $\gamma = 0.97$, it is a saddle point, with stable eigenvectors projected onto the diagonal of the policy space and unstable eigenvectors directed perpendicular to it. 
The trajectory initialised at $\pi^i_C(0) = 0.9$ remains at mutual cooperation $(\pi^i_C \approx 1)$ within any finite number of steps feasible for computational simulation. Note however that the equations show that this is \emph{not} a true fixed point and pure policies are prohibited due to $T>0$.
}
\label{fig:5}
\end{figure}

\subsection{Stability Analysis}
% -------------- Fixed point -------------- 
The four-dimensional fixed point $\mathbf Q_*$ of \eqref{eq:QLmodelNEW} is obtained by finding the roots of the second term for all $i, a^i$. The coupled equations read
\begin{equation}
    \begin{aligned}
    \label{eq:FixedPointQ}
    Q^{i}_{a^i *}
    &:= 
    \mathbb{E}_{A^{-i} \sim \pi^{-i}_*}  R^i_{a^i A^{-i}} 
    + \gamma \max_{b^i \in \mathcal A^i} Q^i_{b^i *} \\
    &= \mathbb{E}_{A^{-i} \sim \pi^{-i}_*}  R^i_{a^i A^{-i}} 
    + \gamma \max_{b^i \in \mathcal A^i}
    \sum_{k=0}^\infty \gamma^k \mathbb E_{A^{-i} \sim \pi^{-i}_*}  R^i_{b^i A^{-i}} \\
    &= \mathbb{E}_{A^{-i} \sim \pi^{-i}_*}  R^i_{a^i A^{-i}} 
    + 
    \underbrace{
    \frac{\gamma}{1 - \gamma} 
    \max_{b^i \in \mathcal A^i}
    \mathbb E_{A^{-i} \sim \pi^{-i}_*} R^i_{b^i A^{-i}}
    }
    _{\text{constant in } a^i}
    .
    \end{aligned}
\end{equation}
Note that in the translation of $\mathbf Q_*$ to $\boldsymbol \pi_*$ via \eqref{eq:BoltzmannPolicy}, the second term of \eqref{eq:FixedPointQ} is irrelevant as it is an offset constant in $a^i$ and only the differences of the $Q$-values matter. This means that a fixed point of the dynamics described by \eqref{eq:deterministic policy update} is also a fixed point of \eqref{eq:QLmodelNEW} in \emph{policy space}, and vice versa. So why does the new model behave so differently?

The key lies in \emph{stability}. Although both models share the same unique fixed point in policy space, their stability properties differ. While it is a stable node for all values of $T$ and all values of $\gamma$ in the BQL and FAQ model, it is more nuanced in the new model.
For $T = 1$, a linear stability analysis (figure \ref{fig:4}, appendix \ref{sec:Appendix_QL}) reveals that the fixed point is a stable focus attractor for $\gamma \lesssim 0.75$, meaning that eventually all trajectories converge to the fixed point. But at $\gamma_{cr_1} \approx 0.75$, the system undergoes a supercritical Neimark-Sacker bifurcation\footnote{A Neimark-Sacker bifurcation is the discrete-time equivalent of an Andronov-Hopf bifurcation.}. This turns the stable focus into an unstable focus, around which a stable limit cycle emerges. All trajectories with asymmetric initial conditions in policy space, even with minimal deviation, converge to the limit cycle instead of the fixed point. This describes the oscillations observed for $\gamma = 0.8$ in figure \ref{fig:1} and \ref{fig:2}. For $\gamma \gtrsim 0.95$, the unstable focus turns into a saddle node.

Figure \ref{fig:5} depicts these different dynamical regimes by plotting projections into 2D policy space and a constructed 3D space, defined by the basis vectors $\mathbf{q}_1 = (1, -1, 0, 0)$, $\mathbf{q}_2 = (0, 0, 1, -1)$, and $\mathbf{q}_3 = (1, 1, -1, -1)$.  
The first two dimensions are the $\Delta Q^i$-values, the third dimension indicates difference between agents.
Note that the trajectory initialised at $\pi^i_C(0) = 0.9$ for $\gamma = 0.97$ remains at mutual cooperation $(\pi^i_C \approx 1)$ within any finite number of steps feasible for computational simulation. However, the equations show that this is \emph{not} a true fixed point.\footnote{Technically, $\pi^i_C = 1$ would be a fixed point, but any finite $T > 0$ prohibits pure policies.} 

So far we limited the discussion to $T=1$. As noted in section \ref{sec:Discrepancies} (figure \ref{fig:2}), the position of the fixed point in policy space changes with varying $T$. The stability analysis (figure \ref{fig:4}) further reveals that the effect of the discount factor $\gamma$ also varies for different $T$. 

% -------- 
\subsection{Cause of Metastable Phases and Oscillations} 
With the deterministic equation \eqref{eq:FixedPointQ} established for calculating an agent's target values based on its opponent's policy, we now examine the underlying causes of the metastable phases and oscillations observed in figure \ref{fig:1}, which are matched by our model. 
Our discussion focuses on the trajectory starting from the initial policy $(\pi^1_C(0), \pi^2_C(0)) = (0.5, 0.48)$, though similar reasoning applies to all other initial conditions.
Note that our model cannot precisely capture the exact timing of specific runs due to the inherent randomness and sensitivity to initial actions, but it effectively captures the overall timescales of the stochastic system. 

\paragraph{Metastable Phases}
    Starting at zero, all $Q$-values grow. Since defection yields higher rewards than cooperation, the growth rate of $Q^i_D$ is higher than of $Q^i_C$. This in turn causes the difference $\Delta Q^i = Q^i_D - Q^i_C$ to increase, resulting in a fast decline of the probability to cooperate. 
    The policy of the second agent declines slightly faster than the first, approaching $\pi^2_C \approx 0$. At this point, given $\pi^2_C$, Agent 1's corresponding target values, calculated using \eqref{eq:FixedPointQ}, are
    \begin{align*}
        Q^1_{C, target} = 0 + \frac{\gamma}{1 - \gamma} = 4, \\
        Q^1_{D, target} = 1 + \frac{\gamma}{1 - \gamma} = 5,
    \end{align*}
    resulting in $\pi^1_C \approx 0.27$. In return, given $\pi^1_C$, Agent 2's target values are $Q^2_{C, target} \approx 9.1$ and $Q^2_{D, target} \approx 10.4$. 
    
    Since Agent 2 primarily defects, $Q^2_D$ updates frequently and reaches its target quickly, while $Q^2_C$ lags due to infrequent updates, keeping $\pi^2_C$ near zero. This metastable phase persists until $Q^2_C$ receives enough updates to approach its target. Over time, $Q^2_C$ gradually catches up, closing the gap $\Delta Q^2$, and the assumption $\pi^2_C \approx 0$ no longer holds.
    
\paragraph{Oscillations}
    As $\pi^2_C$ grows, the expected rewards and hence also the target values of agent $1$ grow. But again, due to the asymmetric update frequency, $Q^1_D$ increases much faster than $Q^1_C$. The policy $\pi^1_C$ plummets close to zero. 
    This has the effect that the target values of agent $2$ decrease drastically as well, closing the $\Delta Q^2$ gap even further. As a result, $\pi^2_C$ grows rapidly. Now, the roles of agent $1$ and $2$ are swapped and the process begins all over again, albeit with a shorter period. An oscillating pattern emerges.
    The oscillations can be understood as a feedback loop in which the agents' adaptations consistently lag behind the changes of their effective environment. This phenomenon, known as the 'moving target problem' in RL \cite{Sutton1998},  poses a significant challenge in MARL \cite{marl-book, hernandez2017survey}.

\section{Discussion and Conclusion} \label{sec:Discussion}
% - We introduced a corrected model of independent QL
Our analysis underscores the importance of accounting for $Q$-value update frequencies to understand independent Q-learning dynamics. 
By incorporating these frequencies, our deterministic approximation captures behaviours that simpler policy-space approximations like the BQL/FAQL model cannot describe. 
This distinction becomes particularly evident in the example of the Prisoner's Dilemma, where we have shown that the resulting 4D dynamics 
not only exhibit different transient dynamics than the 2D FAQL/BQL model, but can also prevent convergence to a joint policy, by altering the stability properties of equilibria.
It is therefore crucial to recognise that the FAQL and BQL models do not represent classical, incremental Q-learning but rather specifically modified variants---a non-trivial nuance sometimes overlooked in the literature. 
While our focus in this work has been on single-state environments as a minimalistic case study, the insights gained are also relevant to multi-state environments, particularly those with recurring state transitions. For instance, in environments with two states and very low transition probabilities between them, similar independent Q-learning behaviour would likely emerge.

% - metastable phases
Our case study illustrates how using a Boltzmann policy during independent Q-learning can induce metastable phases by causing update frequencies to approach zero. The time required for the corresponding $Q$-value to reach its target can far exceed any realistic number of learning steps. These metastable phases could therefore easily be mistaken for equilibrium dynamics, posing a risk of misinterpretation. This highlights the importance of examining all dynamic variables (e.g., all $Q$-values) rather than focusing solely on the target variables of interest (e.g., the policy), as only a few of these might display perceptible drift during a metastable phase that indicates the instability \cite{kittel2017timing}.
% - Highlighting transient cooperation
We demonstrated this issue by addressing the question under what conditions independent incremental algorithms spontaneously ``learn" to cooperate in social dilemmas \cite{BarfussMeylahn2023}. Specifically, we showed that what might initially appear as stable cooperative behaviour in the Prisoner's Dilemma—seemingly contradicting the rationale of strategic interactions—is, in fact, a prolonged transient phase of the Q-learning process rather than a true equilibrium.
While such misinterpretations are relatively easy to avoid in simple environments like the Prisoner's Dilemma, they become far more challenging in complex environments with many agents, actions, and multiple Nash equilibria. 
%In this regard, a complementary Dynamical Systems perspective can be helpful.

% - moving target problem
Furthermore, we showed how the moving target problem can cause stable oscillations, which prevent convergence to a joint policy. 
In our case study, these phenomena are tied with higher values of $\gamma$, which induce a Neimark-Sacker bifurcation.
Nevertheless, the moving target problem is not unique to scenarios with $\gamma > 0$. In other settings, such as a public goods game with multiple agents where $\gamma = 0$, different mechanisms can similarly intensify this issue, resulting in comparable phenomena.
% --- Root Cause: Non-Stationarity ---
Although algorithmic adjustments such as batch learning, frequency-adjusted updates, or adopting alternative policy mechanisms (e.g. epsilon-greedy) can help mitigate the moving target problem, these are, essentially, \emph{symptomatic} treatments. 
The underlying root cause of oscillatory or even more complex behaviour lies in the \textit{non-stationarity} of the effective environment for each agent, a fundamental challenge in MARL \cite{hernandez2019survey}.

% - rather a bug than a feature
If the independent Q-learning algorithm is interpreted as a model of actual learning processes occurring in humans or other organisms, the described complex dynamics should be considered interesting features worthy of further study. Most of the time, however, MARL algorithms are not meant as a model of something but as a numerical tool for finding certain types of strategic equilibria (such as Quantal Response Equilibria). For that application, the described complex dynamics should rather be considered a bug than a feature as it makes the MARL tool less useful.
In that context, addressing the non-stationarity challenge is crucial for developing scalable MARL algorithms with robust convergence guarantees, which remains an open research problem \cite{marl-book}. 
As demonstrated, a Dynamical Systems perspective can be helpful for future work in this regard. 

% - Future directions
Looking ahead, future research could focus on extending our approximate model to multi-state environments, partially observable stochastic games, joint-action learning and other Temporal-Difference learning algorithms, broadening its applicability to more complex settings. Additionally, incorporating a noise term into the deterministic ordinary difference equations to create stochastic difference equations would provide a more accurate representation of the inherent stochasticity in these algorithms, e.g., by making better predictions about exit times from metastable phases and average periods of oscillations.

% ---------------------- Suppl. Info, References ---------------------- 

% \paragraph{Code availability.} All computer code will be made publicly available upon acceptance of this manuscript.

\paragraph{Acknowledgments.} W.B. acknowledges support from the Cooperative AI Foundation.

\section*{Appendix}

\appendix
\label{appendix}

% -------------- Section --------------
\section{Q-Learning} \label{sec:Appendix_QL_original}
The original \emph{single-agent} incremental Q-learning algorithm \citep{watkins1992} is defined in the framework of a finite Markov Decision Processes (MDP) \cite{marl-book}, consisting of a finite non-empty set of states $\mathcal S$, a subset of terminal states $\mathcal S_{terminal} \subset \mathcal S$, a finite non-empty set of actions $\mathcal A$, a reward function $R: \mathcal{S} \times \mathcal{A} \times \mathcal{S} \rightarrow \mathbb{R}$ and a state transition probability function $ T: \mathcal{S} \times \mathcal{A} \times \mathcal{S} \rightarrow [0,1] $, such that for all $s \in \mathcal{S}, a \in \mathcal{A}: \sum_{s' \in \mathcal{S}} T(s'| a, s) = 1$.

At each time step $t$, a singular Q-learning agent observes state $S(t)=s$ of the environment, chooses action $A(t)=a$, upon which the environment transitions to state $S(t+1)=s_{next}$ and the agent receives the reward $R(s,a,s_{next}) = r$. The agent then updates its value estimate of the state-action pair $(s,a)$, called $Q$-value, via the update rule
\begin{align}
    \label{eq:original_Q-learning_update_rule}
    Q_{s,a}(t+1)
    &= 
    Q_{s, a}(t) + 
    \alpha
    \left [
        r + \gamma \max_{b \in \mathcal A} Q_{s_{next}, b}(t) - Q_{s, a}(t) 
    \right ]
    ,
    \\
    Q_{s',a'}(t+1)
    &= 
    Q_{s', a'}(t)
    \qquad \text{for all } (s',a')\neq (s,a)
    ,
\end{align}
where $\alpha \in [0,1)$ is the agent's \textit{learning rate}, and the \textit{discount factor} $\gamma \in [0,1)$ determines the weight the agent assigns to the current estimate of the optimal value of the next state $s_{next}$. Note that \emph{only} the $Q$-value of the state-action pair actually played at time $t$ gets updated, the remaining $Q$-values retain their current values. 
% Q-learning is guaranteed to converge to optimal state-action values under certain conditions, which includes among other that the environment is stationary \cite{watkins1992}.
Q-learning is guaranteed to converge to optimal state-action values under certain conditions \cite{watkins1992}, with one key requirement being that the environment remains stationary---a crucial property of an MDP.

% -------------- Section --------------
\section{Derivation of the BQL Model} \label{sec:Appendix_BQL_derivation}
In the limit $K \rightarrow \infty$ (and subsequently $K_{a^i} \rightarrow \infty$), the stochastic batch temporal difference error \eqref{eq:batchTemporalDifferenceError} becomes almost surely (a.s.) deterministic because of the law of large numbers. It can be written in dependence of all $Q$-values at time t as
\begin{equation}
    \begin{aligned}
    \label{eq:batchTemporalDifferenceErrorDeterministic}
    D^i_{a^i, \mathbf Q (t)}  
        &:= 
        \lim_{K \rightarrow \infty}
        D^i_{a^i, \mathbf A(t), \dots ,\mathbf A(t + K), Q^i(t)} 
        \\
        &\overset{\text{a.s.}}{=}
        \mathbb E_{A^i(t) = a^i, A^{-i}(t) \sim \pi^{-i}(t)}
        \left (
            \delta_{A^i(t) a^i}
            \left[ 
                R^i_{\mathbf A (t)}
                + \gamma^i \max_{b^i \in \mathcal A^i} Q^i_{b^i}(t) 
                - Q^i_{a^i}(t) 
            \right]
        \right )
        \\
        % &= 
        % \mathbb E_{A^{-i}(t) \sim \pi^{-i}(t)}
        % R^i_{a^i A^{-i}(t)}
        % + \gamma^i \max_{b^i \in \mathcal A^i} Q^i_{b^i}(t) 
        % - Q^i_{a^i}(t) 
        % \\
        &=
        \mathbb E 
        _{A^{-i}(t) \sim \pi^{-i}(t)}
        R^i_{a^i A^{-i}(t)} 
        + \underbrace{
        \gamma^i
        \max_{b^i\in \mathcal A^i} 
        Q^i_{b^i}(t) 
        }_{\textrm{constant in $a^i$}} 
        \\
        &\quad - T^i \ln \pi^i_{a^i}(t)  
        - \underbrace{
        T^i \ln \sum_{b^i \in \mathcal A^i} 
        \exp[ Q^i_{b^i}(t) /T^i ] 
        }_{\textrm{constant in $a^i$}}
        ,
    \end{aligned}
\end{equation}
where the last two terms are the inverse of \eqref{eq:BoltzmannPolicy}.
% Note that the term $ \gamma^i \max_{b^i\in \mathcal A^i} Q^i_{b^i}(t)$ does not depend on $a^i$ because no matter which actions the agents choose, the agents assume that the environment transitions back to the same state with probability $\gamma^i$. 
The deterministic update rule for the $Q$-values in the separated \emph{update} timescale $u$ then reads
\begin{equation}
\label{eq:Q-learning_updateRule_QBL_limit}
    Q^i_{a^i}(u+1) 
    = 
    Q^i_{a^i}(u) + 
    \alpha^i
    D^i_{a^i, \mathbf Q (u)}
    .
\end{equation}
Inserting \eqref{eq:Q-learning_updateRule_QBL_limit} into \eqref{eq:BoltzmannPolicy} returns a deterministic update rule for the policy, 
\begin{equation}
\begin{aligned}
\label{eq:deterministic policy updatePRE}
    \pi^i_{a^i}(u + 1)
    &= \frac{
        \exp[ 
            Q^i_{a^i}(u+1)  / T^i
            ]
        }{
        \sum_{b^i \in \mathcal A^i} 
        \exp[ 
            Q^i_{b^i}(u+1)  / T^i]
        } \\
    &= 
    \frac{
    \exp[ Q^i_{a^i}(u) / T^i ]
    \exp[ \alpha^i
        D^i_{a^i, \mathbf Q (u)}  / T^i ]
    }{
    \sum_{b^i \in \mathcal A^i} 
    \exp[ Q^i_{b^i}(u) / T^i ]
    \exp[ \alpha^i
        D^i_{b^i, \mathbf Q (u)}  / T^i ]
    }
    \\
    &= 
    \frac{
    \pi^i_{a^i}(u) 
    \exp[ \alpha^i
        D^i_{a^i, \mathbf Q (u)}  / T^i ]
    }{
    \sum_{b^i \in \mathcal A^i} \pi^i_{b^i}(u) 
    \exp[ \alpha^i  
    D^i_{b^i, \mathbf Q (u)}  / T^i ]
    }
    .
\end{aligned}
\end{equation}
As it is, \eqref{eq:deterministic policy updatePRE} depends on the four-dimensional vector $\mathbf Q (u)$. To have an approximation that conveniently reduces the learning dynamics to the two-dimensional policy space, \eqref{eq:deterministic policy updatePRE} needs to be expressed purely in terms of $\boldsymbol \pi (u)$.
Luckily, one can make use of the fact that \eqref{eq:deterministic policy updatePRE} is invariant under adding terms to $D^i_{a^i, \mathbf Q (u)}$ that are constant in $a^i$, such as the last term of \eqref{eq:batchTemporalDifferenceErrorDeterministic}. Note that in single-state environments, also the second term including the discount factor is constant in action---no matter which actions the agents choose, the environment transitions back to the same unique non-terminal state---and can thus be excluded. This means that the dynamics are \emph{independent} of the discount factor.
Equation \eqref{eq:deterministic policy updatePRE} simplifies to
\begin{equation}
\begin{aligned}
\label{eq:deterministic policy update}
    \pi^i_{a^i}(u + 1)
    &= \frac{
    \pi^i_{a^i}(u) 
    \exp[ \alpha^i
        D^i_{a^i, \boldsymbol \pi(u)}  / T^i ]
    }{
    \sum_{b^i \in \mathcal A^i} \pi^i_{b^i}(u) 
    \exp[ \alpha^i  
    D^i_{b^i, \boldsymbol \pi(u)}  / T^i ]
    }
    ,
\end{aligned}
\end{equation}
where
\begin{equation}
\begin{aligned}
\label{eq:ReducedDeterministicTDerror}
    D^i_{a^i, \boldsymbol \pi(u)} 
        &:=
        \mathbb E _{A^{-i}(u) \sim \pi^{-i}(u)}
        R^i_{a^i A^{-i}(u)} 
        - T^i \ln \pi^i_{a^i}(u)  
        .
\end{aligned}
\end{equation}

% -------------- Section --------------
\section{Stability Analysis of the BQL Model}\label{sec:Appendix_BQL}
% The fixed point policy $\boldsymbol \pi_{*}$ of the BQL model \eqref{eq:deterministic policy update} is defined by $\boldsymbol \pi_{*}(u+1) = \boldsymbol \pi_{*}(u)$. It can be determined by finding the roots of \eqref{eq:ReducedDeterministicTDerror} for all $i, a^i$. 
% After normalisation, this results in the two dimensional system of equations
% \begin{equation}
% \label{eq:FixedPointPi_2}
%     \pi^i_{C*} = \frac{
%     \exp[ 
%         \mathbb E _{A^{-i} \sim \pi^{-i}_*}  R^i_{C A^{-i}} /T  
%         ]
%     }{
%     \sum_{b^i \in \mathcal A^i} 
%     \exp[ 
%         \mathbb E _{A^{-i} \sim \pi^{-i}_*}  R^i_{b^i A^{-i}} /T 
%         ]
%     } 
%     .
% \end{equation}
We solve the two-dimensional system of equations \eqref{eq:FixedPointPi} numerically using the \texttt{fsolve} function from Python's SciPy library.
For the Prisoner's Dilemma, there exists a unique symmetric fixed point (see figure \ref{fig:2}).
To determine its stability, we conduct a linear stability analysis at the fixed point. To this end, we calculate the Jacobian,
% \todo{Jobst: warum ist die Diagonale null und nicht links oben von der Form $pq [\pi^2_C(R_{CC}-R_{DC}) + \pi^2_D(R_{CD}-R_{DD})] / T[p+q]^2$?}
% Answer: Weil Gleichung \eqref{eq:FixedPointPi} nur von der policy des Gegners abhängt, nicht aber von der eigenen policy. Daher ist die partielle Ableitung null.

\begin{equation}
\label{eq:JacobiMatrixPi}
J 
=
    \left (
    \begin{array}{cc}
        \partial_{\pi^1_C} \pi^1_C & \partial_{\pi^2_C} \pi^1_C \\
        \partial_{\pi^1_C} \pi^2_C & \partial_{\pi^2_C} \pi^2_C
    \end{array}
    \right )
= 
\left (
    \begin{array}{cc}
        0 
        & - \frac{p^1_{\pi^2} q^1_{\pi^2}}{T[p^1_{\pi^2} + q^1_{\pi^2}]^2} 
        \\
        - \frac{p^2_{\pi^1} q^2_{\pi^1}}{T[p^2_{\pi^1} + q^2_{\pi^1}]^2} 
        & 0
    \end{array}
    \right )
    ,
\end{equation}
where
\begin{align*}
    p^i_{\pi^{-i}}
    &:= \exp[ \mathbb E_{A^{-i} \sim \pi^{-i}} R^i_{a^i=C, A^{-i}} / T] 
    ,
    \\
    q^i_{\pi^{-i}}
    &:= \exp[ \mathbb E_{A^{-i} \sim \pi^{-i}} R^i_{a^i=D, A^{-i}} / T] 
    .
\end{align*}
Note that the prefactor $-1$ in \eqref{eq:JacobiMatrixPi} comes from
\begin{equation*}
    R^i_{a^i=C, a^{-i}=C} - R^i_{a^i=C, a^{-i}=D} - R^i_{a^i=D, a^{-i}=C} + R^i_{a^i=D, a^{-i}=D} = 3 - 0 - 5 + 1 = -1
\end{equation*}
We calculate the Eigenvalues $\lambda_n$ of the Jacobi matrix numerically with the function \texttt{numpy.linalg.eig} from Python's NumPy library.
Since all eigenvalues are $|\lambda_n| < 1$, we deduct the discrete-time fixed point to be a stable node.

% -------------- Section --------------
\section{Stability Analysis of our Model}\label{sec:Appendix_QL}
% To analyse the stability of the four dimensional fixed point $\mathbf Q_* = (Q^1_{C*}, Q^1_{D*}, Q^2_{C*}, Q^2_{D*})$, we need to analyse the eigenvalues and eigenvectors of the Jacobi matrix at this point. 
If we take into account that for the Prisoner's Dilemma, $Q^i_{C*} < Q^i_{D*}$ holds at the fixed point $\mathbf Q_*$, the maximum term of \eqref{eq:QLmodelNEW} reduces to $\max(Q^i_{C*}  Q^i_{D*})= Q^i_{D*}$. We can therefore simplify \eqref{eq:QLmodelNEW} at the fixed point $\mathbf Q_*$ to
\begin{equation}
\label{eq:QLmodelNEW_appendix}
    \begin{aligned}
        \mathbb E_{ \mathbf A(t) \sim \boldsymbol \pi(t) } [ Q^i_{a^i}(t+1) \mid Q^i_*(t) ] 
        &= 
        Q^i_{a^i*}(t)
        \\ 
        &\quad 
        + \alpha
        \pi^i_{a^i*}(t) 
        \biggl [
        \mathbb E_{ A^{-i}(t) \sim \pi^{-i}_*(t) }
        R^i_{a^i A^{-i}(t)}
        + \gamma Q^i_{D*}
        - Q^i_{a^i*}
        \biggr ]
        .
    \end{aligned}
\end{equation} 
To shorten the notation, we omit the dependencies and the fixed point subscript index $*$ in the following, and make use of the relations
\begin{align*}
    \partial_{Q^i_C} \pi^i_C &= \partial_{Q^i_D} \pi^i_D = \frac{ e^{(Q^i_C + Q^i_D)/T} }{ T(e^{Q^i_C/T}+e^{Q^i_D/T})^2 } , \\
    \partial_{Q^i_D} \pi^i_C &= - \partial_{Q^i_C} \pi^i_C = - \partial_{Q^i_D} \pi^i_D = \partial_{Q^i_C} \pi^i_D. 
\end{align*}
To shorten the notation further, we introduce
\begin{align*}
    f^i &:= 
        \alpha 
        \partial_{Q^i_C}  \pi^i_C 
        \left [
        \pi^{-i}_C R^i_{a^i=C, a^{-i}=C} + 
        (1 - \pi^{-i}_C) R^i_{a^i=C, a^{-i}=D}
        + \gamma Q^i_D
        - Q^i_C 
        \right] 
        , \\
    g^i &:= 
        \alpha 
        \pi^i_C 
        \partial_{Q^{-i}_C}  \pi^{-i}_C
        \left [
        R^i_{a^i=C, a^{-i}=C} -
        R^i_{a^i=C, a^{-i}=D}
        \right] 
        , \\
    h^i &:= 
        \alpha 
        \partial_{Q^i_C}  \pi^i_C 
        \left [
        \pi^{-i}_C R^i_{a^i=D, a^{-i}=C} + 
        (1 - \pi^{-i}_C) R^i_{a^i=D, a^{-i}=D}
        - (1 - \gamma) Q^i_D
        \right] 
        , \\
    k^i &:= 
        \alpha 
        (1 - \pi^i_C )
        \partial_{Q^{-i}_C}  \pi^{-i}_C
        \left [
        R^i_{a^i=D, a^{-i}=C} -
        R^i_{a^i=D, a^{-i}=D}
        \right]
        ,
\end{align*}
which help to write the Jacobi matrix at the fixed point as
\begin{equation*}
J =
    \left (
    \begin{array}{llll}
        f^i - \alpha \pi^i_C + 1
        & - f^i + \alpha \gamma \pi^i_C
        & g^i
        & - g^i 
        \\
        - h^i 
        & h^i - \alpha (1 - \gamma) (1 - \pi^i_C) + 1
        & k^i 
        & - k^i 
        \\
        g^{-i}
        & - g^{-i} 
        & f^{-i} - \alpha \pi^{-i}_C + 1
        & - f^{-i} + \alpha \gamma \pi^{-i}_C
        \\
        k^{-i} 
        & - k^{-i} 
        & - h^{-i} 
        & h^{-i} - \alpha (1 - \gamma) (1 - \pi^{-i}_C) + 1
        \\
    \end{array}
    \right )
    .
\end{equation*}
We solve the eigenvalues of the Jacobi matrix at the fixed point \eqref{eq:FixedPointQ} numerically with the function \texttt{numpy.linalg.eig} from Python's NumPy library.
The absolute eigenvalues are plotted against the discount factor in figure \ref{fig:5} for three different temperature values.

% ---------------------- Bibliography ---------------------- 
% Choose bibliography style above at the beginning: unsrt: usorted, order of appearances
\bibliography{bibliography} % Specify your .bib file

\end{document}